\documentclass[10pt]{article}
\usepackage[letterpaper,margin=1in]{geometry}
\usepackage[utf8]{inputenc}
\usepackage[T1]{fontenc}
\usepackage{bm}
\usepackage[most]{tcolorbox}
\usepackage{xcolor}
\usepackage{hyperref}
\usepackage{ragged2e}
\usepackage{titlesec}
\usepackage{amssymb}
\usepackage[numbers,sort&compress]{natbib}
\usepackage{authblk}

\providecommand{\JournalTitle}[1]{#1}

\usepackage{graphicx}
\usepackage{wrapfig}

\usepackage{floatflt}
\definecolor{boxcolor}{HTML}{dcf0fa}
\definecolor{framecolor}{HTML}{cce0eb}
\tcbset{
  naturebox1/.style={
    colback=boxcolor,       % light gray background
    colframe=framecolor,        % black frame
    coltext=black,
    coltitle=black,
    boxrule=0.4pt,         % thin border
    arc=0.5mm,             % slight rounding
    left=2mm, right=2mm, top=2mm, bottom=2mm,
    fonttitle=\bfseries,
    fontupper=\small,
    title={Box 1 \textbar\ Properties of Solid-State Quantum Emitters},
    boxed title style={
      top=1.5mm, bottom=1.5mm  % << small vertical padding
    },
  }
}
\tcbset{
  naturebox2/.style={
    colback=boxcolor,       % light gray background
    colframe=framecolor,        % black frame
    coltext=black,
    coltitle=black,
    boxrule=0.4pt,         % thin border
    arc=0.5mm,             % slight rounding
    left=2mm, right=2mm, top=2mm, bottom=2mm,
    fonttitle=\bfseries,
    fontupper=\small,
    title={Box 2 \textbar Emitter Interactions},
    boxed title style={
      top=1.5mm, bottom=1.5mm  % << small vertical padding
    },
  }
}

\tcbset{
  naturebox3/.style={
    colback=boxcolor,       % light gray background
    colframe=framecolor,        % black frame
    coltext=black,
    coltitle=black,
    boxrule=0.4pt,         % thin border
    arc=0.5mm,             % slight rounding
    left=2mm, right=2mm, top=2mm, bottom=2mm,
    fonttitle=\bfseries,
    title={Box 3 \textbar Photon Mediated Interactions},
    boxed title style={
      top=1.5mm, bottom=1.5mm  % << small vertical padding
    },
  }
}

\definecolor{hotpink}{RGB}{255, 105, 180}
\definecolor{darkred}{RGB}{214, 39, 40}

\renewenvironment{abstract}
	{\quotation}
	{\endquotation}

\title{Many-Body Entanglement in Solid-State Emitters} 
\date{}
\author{}
\begin{document}
\raggedbottom
\maketitle
\begin{center}
\author{Emma Daggett}$^1$,
\author{Christian M. Lange}$^2$,
\author{Bennet Windt}$^{3,4}$,
\author{Arshag Danageozian}$^{5,6}$,
\author{Alexander Senichev}$^{7,8}$,
\author{Jordi Arnau Montañà-López}$^{3,4}$,
\author{Chanchal}$^1$,
\author{Kinjol Barua}$^{1,7}$ ,
\author{Xingyu Gao}$^2$,
\author{Zhaoyun Zheng}$^9$,
\author{Vijin Kizhake Veetil}$^{10}$,
\author{Souvik Biswas}$^{11}$,
\author{Jonas M. Peterson}$^1$,
\author{Na Liu}$^1$,
\author{Chuchuan Hong}$^9$,

\author{Teri Odom}$^9$,
\author{Matthew Pelton}$^{10}$,
\author{Tongcang Li}$^{2,7}$,
\author{Jelena Vučković}$^{11}$,
\author{Vladamir Shalaev}$^{7,8}$,
\author{Alexandra Boltasseva}$^{7,8}$,
\author{Sophia E. Economou}$^{5,6}$,
\author{Jonathan D. Hood}$^{1,2}$,
\author{Valentin Walther}$^{1,2}$,
\author{Rahul Trivedi}$^{3,4}$, 
\hspace{10pt}
\author{Libai Huang}$^{1,7*}$

\hspace{10pt}

\small$^1${ Department of Chemistry, Purdue University, West Lafayette, IN 47907, USA}\and

\small$^2${ Department of Physics and Astronomy, Purdue University, West Lafayette, IN 47907, USA}\and

\small$^3${ Max Planck Institute of Quantum Optics, Hans-Kopfermann-Straße 1, 85748 Garching, Germany}\and

\small$^4${ Munich Center for Quantum Science and Technology, Schellingstraße 4, 80799 München, Germany}\and

\small$^5${ Department of Physics, Virginia Tech, Blacksburg, VA 24061, USA}\and

\small$^6${ Virginia Tech Center for Quantum Information Science and Engineering, Blacksburg, VA 24061, USA}\and

\small$^7${ Elmore Family School of Electrical and Computer Engineering, Purdue University, West Lafayette, IN 47907, USA}\and

\small$^8${ Birck Nanotechnology Center, Purdue University, West Lafayette, IN 47907, USA}\and

\small$^9${ Department of Chemistry, Northwestern University, Evanston, IL 60208, USA}\and

\small$^{10}${ Department of Physics, University of Maryland, Baltimore County, Baltimore, MD 21250, USA}\and

\small$^{11}${ Department of Electrical Engineering, Stanford University, Stanford, CA 94305, USA}\and

\small[*]{e-mail: libai-huang@purdue.edu}

\end{center}

\begin{abstract} \bfseries \boldmath 
The preparation and control of quantum states lie at the heart of quantum information science (QIS). Recent advances in solid-state quantum emitters (QEs) and nanophotonics have transformed the landscape of quantum photonic technologies, enabling scalable generation of quantum states of light and matter. A new frontier in solid-state quantum photonics is the engineering of many-body interactions between QEs and photons to achieve robust coherence and controllable many-body entanglement. These entangled states, including photonic graph and cluster states, superradiant emission, and emergent quantum phases, are promising for quantum computation, sensing, and simulation. However, intrinsic inhomogeneities and decoherence in solid-state platforms pose significant challenges to realize such complex entangled states. This review provides an overview of the fundamental many-body interactions and dynamics at the light-matter interfaces of solid-state QEs, and discusses recent advances in mitigating decoherence and harnessing robust many-body coherence.
\end{abstract}

\justifying

\section*{Introduction}

\hspace{10pt}
Harnessing the power of quantum information science (QIS) now depends on building entanglement between many qubits while maintaining their coherence~\cite{kimble2008quantum}. Despite exquisite control of individual qubits, the field faces a critical challenge: generating robust and efficient many-body entanglement at scale. While many foundational demonstrations of quantum information have centered around atoms and ions, solid-state platforms offer distinct advantages for scaling to many qubits through their ability to miniaturize and integrate on-chip~\cite{lodahl2018scaling}. Beyond scalability, solid-state quantum emitters (QEs) open new possibilities by leveraging materials science and chemistry to engineer systems with novel functionality, for example, large dipole moments, strong nonlinear interactions, or accessible spin registers. Such enhanced interactions open up new regimes of correlated dynamics, facilitating the creation of large-scale entangled states. Moreover, solid-state platforms also have natural compatibility with integrated photonics, enabling a full suite of on-chip capabilities such as optical manipulation, photon detection, and electronic control (Fig. 1, left). By tailoring the interactions of solid-state QEs—whether through photonic modes or direct dipole-dipole coupling in hybrid light-matter systems—one can generate a wide range of many-body entangled  states~\cite{turunen2022quantum, heinrich2021quantum, evans2018photon, dousse2010ultrabright, cogan2023deterministic, coste2023high, grim2019scalable}. These include metrological states where signal-to-noise scales with photon number $N$ rather than the classical $\sqrt{N}$ limit, cluster states for measurement-based quantum computation, and states with built-in error correction or resilience to photon loss for entanglement distribution (Fig. 1, right). Looking forward, solid-state platforms with tunable interactions will allow us to explore new classes of photonic states that could transform the practicality of quantum photonic technologies.

Significant advances have been made in solid-state QEs and nanophotonics in recent decades (Examples in Fig.1). Various solid-state systems have been harnessed as QEs, including color centers in wide-bandgap materials, epitaxial and colloidal quantum dots (QDs), molecules, and two-dimensional (2D) materials~\cite{atatüre2018material, Senellart2017,  awschalom2018quantum, obrien2009photonic, he2015single, tomm2021bright,  toninelli2021single, aharonovich2022quantum, Zhai2020, gurioli_droplet_2019, uppu2021quantum, gao_single_2025, li_proximity-induced_2023, senichev2021roomtemp}. In parallel, advancements in nanophotonics have enabled the production and manipulation of light on a chip~\cite{ gonzalez2024light, molesky2018inverse, rodt2021integrated, lukin2020integrated, wang_integrated_2020, pelucchi_potential_2022, senichev2024aln}. Despite these advances, achieving robust many-body entanglement in solid-state emitters remains a central challenge. This challenge arises from intrinsic inhomogeneities and dephasing in QEs, which limit the coherence required for robust and scalable entanglement, and integrating them with high quality photonics. A central open question is how to engineer light-matter interactions and collective emitter dynamics to enable not only generation of more photons, but controllable many-body entangled states that are resilient against disorder and noise, and can thus maintain coherence over meaningful time-scales.

This review focuses on the emerging frontier of many-body effects in solid-state quantum photonics. Specifically, we highlight strategies for designing and controlling collective interactions, both between emitters and between photons, to realize complex quantum states such as graph and cluster states, correlated phases, and superradiant emission. Fig. 1 illustrates key concepts and systems through which emitter-emitter and light-matter interactions can be harnessed to produce controllable entangled many-body states. We begin by addressing coherence and control at the single-emitter level, discussing the fundamental limits of optical and spin coherence as well as strategies for stabilization. We then examine mechanisms for realizing many-body interactions between quantum emitters and explore how coupling to nanophotonic structures can be used to engineer, mediate, and protect these interactions from decoherence. Next, we consider approaches to achieving strong photon nonlinearities and review deterministic protocols for generating complex multi-photon entangled states, such as cluster and graph states. Finally, we conclude with an outlook on applications in quantum sensing, metrology, computation, and simulation, and highlight promising directions for future research.

\begin{figure}
    \centering
    \includegraphics[width=1\linewidth]{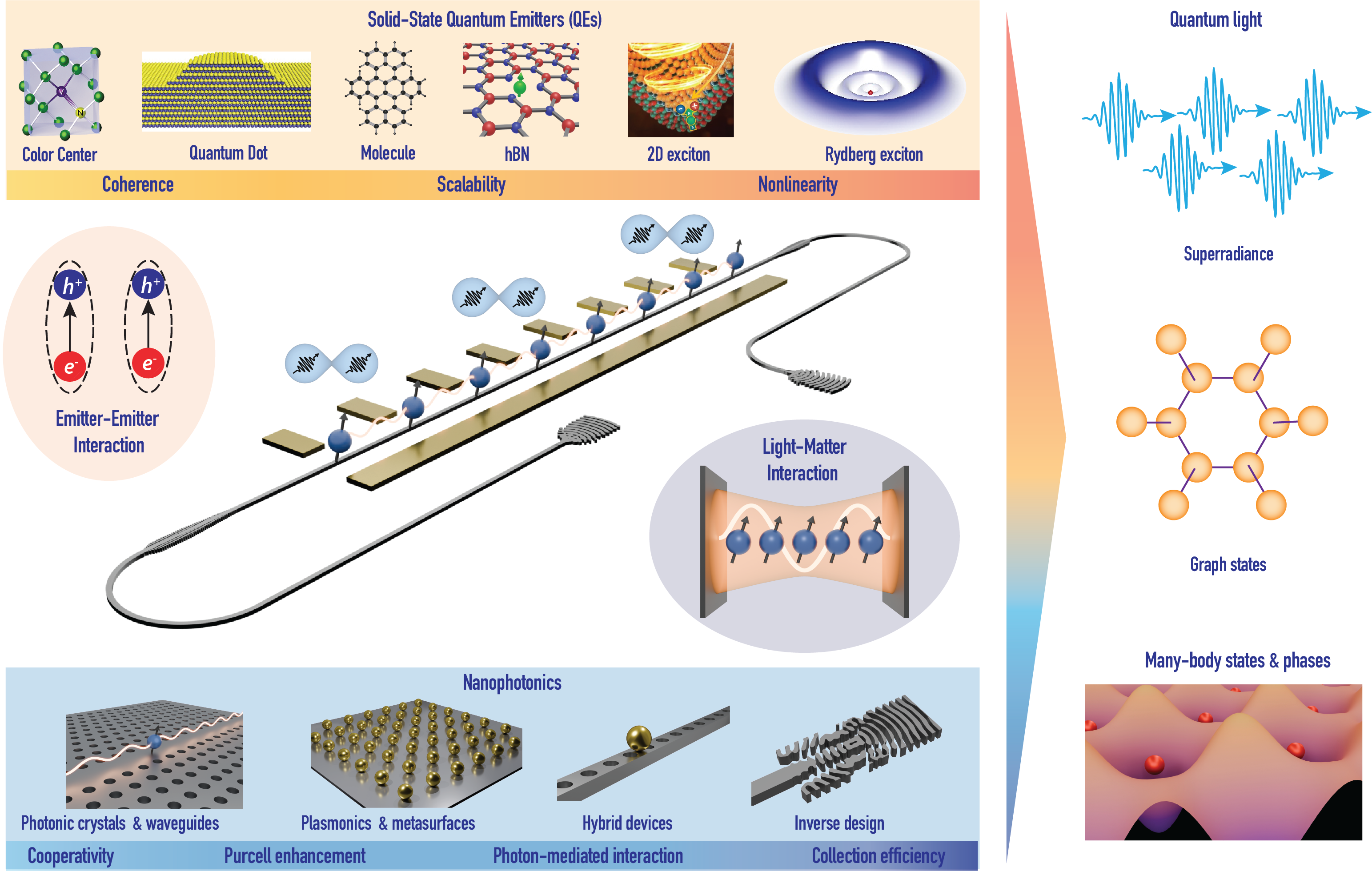}
    \caption{\emph{Overview of achieving many-body entanglement with solid-state quantum emitters integrated in nanophotonics.}
Top: Representative platforms, including vacancies in wide-bandgap crystals, epitaxial quantum dots, molecules, hBN defects, 2D excitons, and Rydberg excitons, highlighting complementary strengths in coherence, scalability, and intrinsic nonlinearity. Quantum dot image reproduced with permission from Ref. \cite{gurioli_droplet_2019}, and image of 2D exciton reproduced with permission from Ref. \cite{li_proximity-induced_2023}. Center: Photonic interfaces enhance light–matter coupling  and mediate long-range emitter–emitter interactions through coupling to shared modes. Bottom: Nanophotonic systems, including photonic crystal cavities and waveguides, plasmonic structures, hybrid devices, and inverse design, enable programmable photon-mediated couplings. Right: These capabilities yield quantum-optical resources (superradiance, multi-photon cluster/graph states, and quantum phase transitions) and enable interacting many-body states.}
    \label{Figure 1}
    
\end{figure}

\section*{Coherence and Control at the Single-Emitter Level}

Solid-state quantum emitters now span several distinct material platforms (Fig.1, top), each offering complementary strengths for quantum photonic applications. Defect centers in wide-bandgap crystals, for example, nitrogen-vacancy (NV) and silicon-vacancy (SiV) centers in diamond, silicon-vacancy (\(V_{\mathrm{Si}}\)) centers in silicon carbide, and emitters in hexagonal boron nitride (hBN), provide stable, atom-like QEs. Epitaxially grown semiconductor QDs and their colloidal counterparts offer strong optical transitions and near-unity quantum efficiencies through bandgap engineering. Two-dimensional materials host tightly bound excitons with large oscillator strengths, enabling room-temperature operation and novel moiré-engineered interactions. Rare-earth ions embedded in crystalline hosts combine narrow optical transitions with exceptionally long-lived spin states. Organic molecules in organic matrices achieve lifetime-limited emission with high photostability. For comprehensive reviews of these various classes of solid-state QEs, readers are directed to recent specialized reviews \cite{aharonovich2016solid,fox2025solidstate, chen2025lowdim}. A major challenge in solid-state QEs is maintaining optical and spin coherence in the presence of the environmental interactions (Box 1). Phonons in the lattice, stray charges, surface effects, nuclear spins, all undermine the ideal two-level character of the system.

\subsection*{Optical coherence}

Photon indistinguishability—having identical frequency, polarization, and temporal profiles—is essential for quantum interference effects: when two indistinguishable single photons arrive simultaneously at a 50:50 beamsplitter, they bosonically coalesce and exit the same port, producing a dip in two-detector coincidences, the Hong-Ou-Mandel (HOM) dip. This indistinguishability enables both interference between consecutive photons from a single emitter and, critically, between photons from multiple emitters to generate entanglement. The degree of indistinguishability directly depends on the emitter's optical coherence.
Box 1 illustrates this cycle of excitation and photon emission, including radiative and nonradiative decay pathways. The coherence of the emitted photon is fundamentally related to the optical transition linewidth, which in turn depends on both population decay time (T$_1$) and total dephasing time (T$_2)$. Since dephasing is typically fast in the solid state due to interactions between the QE and its environment, a key challenge for photon indistinguishability is to achieve a much higher radiative emission rate than the dephasing rate. Near-lifetime limited linewidths (where pure dephasing approaches zero) have been achieved in a few classes of QE systems, including SiV centers in diamond \cite{zuber2023shallow,sipahigil2014indistinguishable}, self-assembled QDs \cite{pedersen2020near,thyrrestrup2018quantum,santori2002indistinguishable}, and organic molecular emitters such as dibenzoterrylene (DBT) \cite{lombardi2021triggered,toninelli2021single}.

The zero-phonon line (ZPL) emission is typically accompanied by broad phonon sidebands at room temperature, evidencing significant phonon coupling. For example, the NV center in diamond has a prominent phonon sideband (only $\approx$ 3\% of its emission decays into the ZPL at room temperature); the SiV center’s inversion-symmetric structure leads to a larger Debye–Waller factor ($\approx$ 70\% of photons decay into the ZPL at cryogenic temperatures) and reduced spectral diffusion. Optical coherence measurements of shallow SiV centers in nano-diamonds have achieved nearly lifetime-limited linewidths of $\sim$100 MHz \cite{zuber2023shallow}. Epitaxial QDs typically have fast radiative rates and can also approach transform-limited emission under resonant excitation at low temperature, promising highly indistinguishable photons \cite{kuhlmann2015transform}, though they remain susceptible to dephasing. In III–V QDs, for example, slow charge noise in the surrounding semiconductor produces spectral diffusion and increases the homogeneous linewidth beyond the transform-limit ($\approx160$ MHz for a 1 ns radiative lifetime), degrading two-photon interference and complicating multi-emitter protocols ~\cite{Kuhlmann2013,Senellart2017}.

In the last decade, a significant amount of research has been conducted on 2D materials such as hBN \cite{ccakan2025quantum,aharonovich2022quantum} and transition metal dichalcogenides (TMDs) \cite{toth2019single}, primarily due to their low cost of fabrication, engineerable optical properties, and relative ease of integration with photonic structures \cite{gao2023atomically,turunen2022quantum,zotev2025nanophotonics}. However, in 2D QEs, coherence times are limited by strong coupling to phonons and susceptibility to spectral diffusion due to local charge fluctuations and strain fields \cite{esmann2024solidstate}. For instance, hBN-based emitters frequently exhibit broad linewidths (several GHz) at room temperature, largely attributed to phonon sidebands and the fluctuating electrostatic environment arising from surface contaminants and intrinsic charge noise \cite{castelletto2020hexagonal,shaik2021optical}. Recent experiments have measured homogeneous linewidths down to tens of MHz at cryogenic temperatures ($\sim$5 K) under resonant excitation \cite{dietrich2018observation}, corresponding to coherence times approaching tens of nanoseconds---although significant emitter-to-emitter variation remains. Similarly, TMD-based quantum emitters, such as localized excitons in $WS_2$, $WSe_2$ monolayers, typically exhibit linewidths broadened by acoustic phonon coupling, spectral wandering, and environmental fluctuations \cite{kianinia2022quantum,he2015single}. Consequently, achieving transform-limited lines in 2D QEs remains a central challenge in obtaining indistinguishable photons. For example, photon indistinguishability from hBN defects remains low – recently, interference visibility of $\approx$~58~\%  was reported at 4~K \cite{fournier2023two, akbari2022lifetime}, whereas a near-unity indistinguishability has been reported for QDs \cite{ding2016demand,somaschi2016near}. 

\subsection*{Coherence at spin-photon interface}
Many quantum emitters, such as the NV center in diamond and the V$_{\text{Si}}$ in silicon carbide, have an optically addressable spin degree of freedom that enables entanglement of multiple spin qubits. These experiments entail the generation of a spin-photon entangled state, followed by the detection of clicks at one detector, heralding the formation of remote spin-spin entanglement. The spin coherence time is generally much longer than the optical coherence time, but it too is limited by environmental noise. Spin dephasing arises largely from magnetic noise coupling to the emitter’s spin, such as interactions with a nuclear and undesired electronic spin bath, spin–orbit coupling to lattice vibrations, or fluctuations in external magnetic and electric fields (see Box 1 for more details). Single NV center electron spins can maintain coherence for a few milliseconds at room temperature \cite{bar2013solid}. This remarkable baseline is due to diamond’s stiff lattice (weak spin–phonon coupling) and low nuclear-spin density, and it can be further extended using dynamical decoupling sequences as discussed below. The SiV center in diamond offers a contrasting case study; its ground-state spin is strongly coupled to orbital degrees of freedom, making it sensitive to lattice vibrations \cite{hepp_electronic_siv_2014}. At liquid-helium temperatures ($\sim$4 K), SiV spins suffer rapid dephasing, on the order of tens of nanoseconds in natural diamond \cite{pingault2017coherent,rogers2014all}.

Recent discoveries of spin-active defects in hBN \cite{vaidya_quantum_2023,aharonovich2022quantum,gong_coherent_2023,gao_single_2025, xu2023hbn} have introduced spin-based quantum functionalities in this 2D material platform. However, spin coherence times ($T_2^*$) for these defects remain limited by hyperfine interactions with abundant nuclear spins, charge noise, and lattice vibrations. Initial coherence measurements revealed very short spin coherence times ($T_{2}^{*}<100$ ns) at room temperature ~\cite{Haykal2022,Rizzato2023}. In epitaxial semiconductor QDs, spins (typically an electron or hole spin confined in the dot) experience a dense nuclear spin environment that tends to limit coherence. Hyperfine interactions with nuclear spins cause rapid dephasing of an electron spin’s precession, giving a free-induction of only a few nanoseconds \cite{cywinski2009pure}. Continued efforts in nuclear spin bath control (through techniques like nuclear spin polarization or feedback) and material engineering (isotopically purified semiconductors) are aimed at extending QD spin coherence close to the millisecond range \cite{gillard2022harnessing,nguyen2023enhanced}.

\subsection*{Suppression of dephasing}
Recent advances have significantly improved the suppression of phonon-induced dephasing (Box 1) in QEs. One notable approach involves coherent two-color excitation of InAs/GaAs QDs, which effectively decouples the exciton from its phonon bath ~\cite{Vannucci2023}. Another promising strategy involves embedding QEs within phononic bandgap structures. By engineering the host semiconductor into a phononic crystal, a bandgap in the phonon density of states at targeted frequencies can be achieved. If the primary phonon coupling of the quantum emitter, typically in the GHz range for acoustic phonons, aligns with this bandgap, spontaneous emission of these phonons can be significantly suppressed ~\cite{Kuruma2025,MacCabe2020}. Additionally, nanomechanical strain engineering, such as creating a spatially varying strain gradient, could localize phonons away from emitter regions, acting as a phonon "focusing lens" ~\cite{Meesala2018}. Another established approach couples QEs to high-finesse optical cavities, preferentially enhancing ZPL emission, an approach explored further in the Photon-Mediated Many-Body Interactions Section.

In addition to optical techniques, dynamical decoupling and strain engineering have significantly enhanced spin coherence times. For instance, recent experiments applying Carr–Purcell–Meiboom–Gill (CPMG) or XY8 dynamical decoupling sequences to ensembles of negatively charged boron vacancies in hBN demonstrated an 100-fold increase in spin coherence time ~\cite{Rizzato2023,Ramsay2023}. Silicon-vacancy (SiV) centers in diamond represent another example. By applying strain, the orbital degeneracy in SiV centers can be lifted, substantially suppressing phonon-induced decoherence without requiring ultra-low temperatures ~\cite{Sohn2018}. Another alternative, is to look for centers with even stronger spin-orbit coupling that remain robust at elevated temperatures, such as the tin-vacancy (SnV) center in diamond - recently shown to host a high-quality spin-photon interface~\cite{rosenthal:2024}.

\subsection*{Inhomogeneity and spectral tuning}
Inhomogeneous broadening (see Box 1) is a major obstacle that arises from fabrication-induced disorder, strain fields, and fluctuations in the local electrostatic environment. This limits photon indistinguishability in emitter ensembles and complicates the creation of entanglement across multiple emitters. Due to inhomogeneous broadening, spectral tunability of QEs is crucial to enabling many-body entanglement. A widely used approach is electrostatic Stark tuning, implemented with microfabricated diode or gate electrodes that shift an emitter’s transition via the quantum-confined Stark effect. This technique is well established for epitaxial QDs and routinely delivers meV-scale tuning. For instance, single In(Ga)As/GaAs QDs exhibit giant Stark shifts of several meV under applied fields ~\cite{Finley2004,Gerardot2007}, and lateral/vertical field configurations have been used to fine-tune excitonic transitions and even the exciton fine structure ~\cite{Gerardot2007,Hofer2019,Luo2012}. Analogous Stark control has been demonstrated for dibenzoterrylene (DBT) molecules embedded in an anthracene via a permanent laser-induced tuning mechanism~\cite{colautti2020lasertuning}. Integration with electrodes allows precise bidirectional control of molecular frequencies~\cite{Duquennoy2024}.

\clearpage

\begin{tcolorbox}[naturebox1, enhanced, breakable]
 
\textbf{Localized single-photon emitters}: 
In an ideal two-level system, the excited state decay rate depends on frequency ($\omega$) and dipole moment ($d$): $\Gamma \propto  \omega^3 d^2$. However, solid-state environments introduce multiple mechanisms for deviation from this behavior \textbf{(a)}.

\textbf{Zero-Phonon Line (ZPL) and Phonon Sidebands:}
Coupling to lattice vibrations and other degrees of freedom partitions optical emission into distinct spectral components \textbf{(a,b)}. The ZPL represents direct electronic transitions, while competing decay pathways include phonon-assisted transitions that create a red-shifted phonon sideband (PSB). Molecular species may also decay into their vibrational manifold. The Debye-Waller factor quantifies the fraction of emission into the ZPL: $DW = I_{ZPL}/(I_{ZPL} + I_{PSB})$. This ranges from $\sim$3\% for NV centers,  to $\sim$70\% for SiV centers, to >95\% for epitaxial QDs. Low ZPL emission may be mitigated by coupling to optical cavities, where the Purcell effect can preferentially enhance decay into the ZPL.

\begin{wrapfigure}[30]{l}{0.65\textwidth} % [lines]{l/r}{width}
  \vspace{-1.5\baselineskip}              % nudge up if needed
  \centering
  \includegraphics[width=\linewidth]{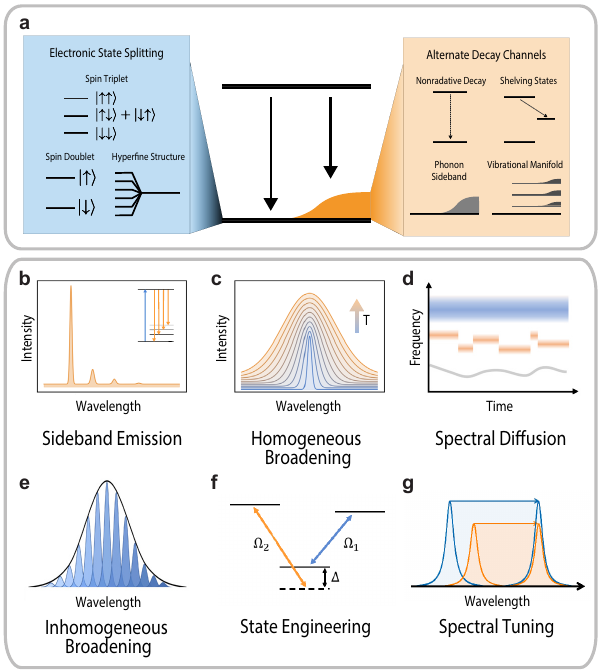}
\end{wrapfigure}

\textbf{Dephasing and spectral wandering:}
Lifetime-limited emission occurs when photon linewidth is determined solely by the emitter's natural lifetime: $\Delta\nu = 1/(2\pi T_1)$. In solid-state systems, interactions with the environment introduce dephasing mechanisms that cause the transition energy to fluctuate, destroying coherence and broadening spectral lines. The optical coherence time $T_2$ is given by $1/T_2 = 1/2T_1+1/T_\phi$, where $1/T_\phi$ is the pure dephasing time. Fast elastic phonon scattering causes homogeneous broadening \textbf{(c)} that increases rapidly with temperature and can be enhanced by low orbital symmetry. Slower processes lead to spectral diffusion \textbf{(d)}: stochastic fluctuations, drift or random-telegraph switching from charge fluctuations, and discrete charge-state conversion or shelving in dark states cause spectral wandering and degrades long-term stability.

\textbf{Inhomogeneous Broadening:}
Solid-state quantum emitters also suffer from inhomogeneous broadening \textbf{(e)}—emitter-to-emitter variations arising from strain fields, local electric fields, and compositional differences that broaden ensemble spectra. Strategies to overcome inhomogeneous broadening include improved sample preparation, post-selection of spectrally matched emitters, and active tuning through electric fields or strain \textbf{(f)} to achieve spectral alignment \textbf{(g)}.

\textbf{Spin coherence}: Many QEs also host an optically addressable spin that can serve as a long-lived quantum memory \cite{burkard2023semiconductor,zhang2020material}. The fine structure of the ground and excited state manifolds is governed by crystal-field symmetry, spin-orbit coupling, and spin-spin interactions. Spin coherence, the ability of a spin to preserve quantum phase relationships over time, is quantified by the longitudinal relaxation time ($T_{1,spin}$) and the transverse coherence time ($T_{2,spin}$) describing the persistence of the spin population and the phase coherence of the superposition states, respectively. Spin dephasing describes the degradation of this coherence, primarily due to magnetic noise from nuclear spin baths, paramagnetic impurities, and spin–phonon processes. Processes such as spin-orbit coupling can dually impact spin and optical coherence.

\textbf{Delocalized emitters}: In addition to localized emitters, excitons in 2D TMDs and other semiconductors can act as quantum emitters whose wavefunctions extend and remain mobile over tens of nanometers to micrometers \cite{azzam2021prospects,gupta2023}. Although typically less coherent, their interactions are long-range and can exhibit van der Waals-type or dipolar scaling depending on excitation configuration. Overlapping excitons enable collective phenomena even without precise positional control \cite{wang2018colloquium}.

\end{tcolorbox}

\section*{Many-Body Interactions, Collective Photon Emission and Quantum Phase Transition}

Achieving entanglement among solid-state QEs requires engineering inter-emitter couplings that can withstand inhomogeneous broadening and environmental dephasing. Solid-state platforms offer a key advantage over cold atoms in this respect; the much higher emitter density can enhance interaction strengths by orders of magnitude, enabling a full hierarchy of collective phenomena, from entangled two-emitter states to mesoscopic ensembles exhibiting superradiant bursts and quantum phase transitions. Box 2 illustrates the key mechanisms for emitter interactions as well as many-body phenomena resulting from these interactions.

\subsection*{Collective emission and entanglement in small emitter arrays (few QEs)}

When the number of emitters is small and each emitter can be individually controlled, one can directly characterize the emergence of entangled states through their collective photon emission. In the simplest case of $N=2$ emitters, dispersive coupling induces a coherent flip-flop interaction between the emitters, under which the single-excitation manifold splits into two entangled eigenstates: a symmetric “bright”/"superradiant" state and an antisymmetric “dark”/"subradiant" state~\cite{zheng2000efficient}. The bright state carries an enhanced dipole moment and radiative rate, whereas destructive interference in the emission pathways of the dark state results in a suppressed decay rate (Box 2).

Spectroscopically, signatures of the entangled states therefore appear as a splitting of the optical resonance peak: the bright state typically shows a broadened, more intense spectral line, while the dark state appears as a narrower, weaker peak, with a splitting between them on the order of $2J$ (Fig. 2a)~\cite{majer_coupling_2007,evans2018photon, lange2024superradiant}. Multidimensional ultrafast spectroscopy has also been used to directly probe multi-exciton collective states, and oscillations in the interaction signal was observed as a function of pulse area, providing clear evidence of coherent exchange dynamics among SiVs in diamond  ~\cite{day2022CoherentSiV}. Alternatively, the modified decay rates of the collective single-excitation states can also be observed directly from time-resolved emission dynamics~\cite{lange2024superradiant,tiranov2023collective}.

A key requirement for the observation of these collective effects is nearly resonant emitters whose coupling $J$ exceeds any difference in transition frequency or linewidth.  Bringing multiple solid-state QEs into resonance is experimentally challenging due to inhomogeneous broadening, as discussed in Box 1 and  Coherence and Control at the Single-Emitter Level section, but recent advances have demonstrated its feasibility. One approach is electrostatic Stark tuning: for example, Trebbia et al.~\cite{trebbia2022tailoring} identified pairs of organic molecules in a crystal and used nanoscale electrodes to finely adjust their transition energies. Another approach, demonstrated by Lange et al.~\cite{lange2024superradiant}, is all-optical tuning: intense laser illumination was used to induce local charge rearrangement and Stark shifts in an organic nanocrystal and bring two embedded molecules into resonance (see Fig.~\ref{fig:superradiance}a-d). In semiconductor systems, highly coherent epitaxial QDs can be tuned into resonance using magnetic fields ~\cite{tiranov2023collective}.

An exciting frontier is controlling larger clusters of individually addressable emitters ($N > 2$) to explore richer many-body quantum dynamics. Already for $N=3$, the Dicke manifold of collective states includes states with multiple excitations delocalized over the ensemble. By dynamically tuning one emitter in and out of resonance with others, one could drive transitions between Dicke states on-demand. This opens the door to realizing a number of theoretical proposals for the tunable emission of multi-photon pulses through controlled coupling between collective bright and dark states~\cite{gonzalez-tudela_deterministic_2015,gonzalez-tudela_efficient_2017,rubies-bigorda_deterministic_2025,abbasgholinejad2025theory}. Although a fully entangled state of $N=3$ solid-state QEs has yet to be observed, the pathway is clear: combine precision tuning (to achieve spectral indistinguishability) with a strong common coupling (via near-field or cavity mediation as described in Box 2) and then use coherent pulse sequences to navigate the multi-excitation state space. Another promising avenue of research is to build complexity by engineering more complicated interactions and adding tunable disorder to the system, while retaining single-emitter control. For instance, few-emitter ensembles have been shown to display intriguing collective emission properties stemming from the coupling to multiple modes and the interplay between spectral and positional disorder~\cite{lukin2023two, lukin_mesoscopic_2025}.

\begin{tcolorbox}[naturebox2, enhanced, breakable]

\begin{wrapfigure}[29]{r}{0.7\textwidth} % [lines]{l/r}{width}
  \vspace{-1.5\baselineskip}              % nudge up if needed
  \centering
  \includegraphics[width=\linewidth]{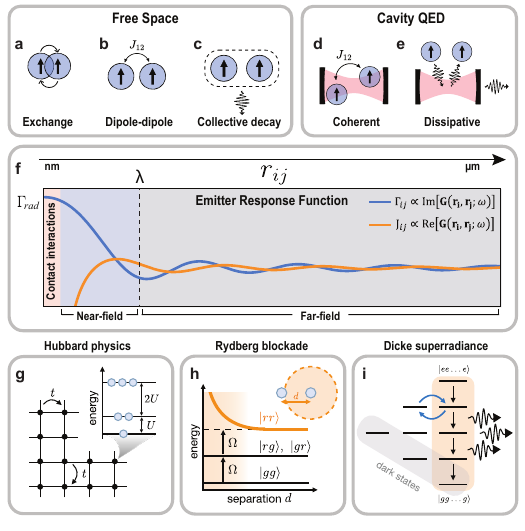}
\end{wrapfigure}

\textbf{Interactions in Free Space}:
Quantum emitters in solids interact through distinct mechanisms depending on their separation. At very short distances below 1 nm, direct wavefunction overlap leads to exchange interactions and Coulomb repulsion \textbf{(a)}. At larger separations, emitters interact through their dipole moments via the electromagnetic field (\textbf{b,c}). At separations less than the transition wavelength, near-field dipole-dipole interactions that scale as $1/r^{3}$ dominate. The far-field radiative coupling between emitters is highly tunable through the electromagnetic environment (\textbf{d,e}), from the extreme confinement of plasmonic structures to the high-Q enhancement of dielectric cavities.

Together, these photon-mediated processes are captured by an effective master equation, where $\rho$ is the density operator and $\sigma^\pm$ are Pauli operators~\cite{lehmberg1970radiation}:%~\cite{gonzalez2024light,jen2025photon}:
\begin{align}
    \frac{d\rho}{dt}
    = -i[H_{\mathrm{eff}},\rho]
    + \sum_{ij} \Gamma_{ij}\left(
        \sigma_i^- \rho \sigma_j^+
        - \frac{1}{2}\{\sigma_i^+\sigma_j^-,\rho\}
    \right),
    \qquad
    H_{\mathrm{eff}} = \sum_{ij} J_{ij}\,\sigma_i^+\sigma_j^- .
\end{align}
Both the coherent $J_{ij}$ and dissipative $\Gamma_{ij}$ coupling strengths are determined by the electromagnetic Green's function $\mathbf{G}(\mathbf{r}_i,\mathbf{r}_j)$, which describes the field at position $\mathbf{r}_i$ due to a dipole source at $\mathbf{r}_j$~\textbf{(f)}.  The coherent coupling is given by the real part  $J_{ij} = \mathrm{Re}[\mathbf{G}(\mathbf{r}_i,\mathbf{r}_j)]$ and enables excitation exchange between emitters through virtual photon exchange \textbf{(b)}. The dissipative coupling is given by the imaginary part $\Gamma_{ij} = \mathrm{Im}[\mathbf{G}(\mathbf{r}_i,\mathbf{r}_j)]$ and describes correlated spontaneous emission through real photon exchange, giving rise to collective superradiant and subradiant states~\textbf{(c)}.

\textbf{Interactions in a cavity}: 
Structuring the electromagnetic field environment allows for the spatial tailoring of the Green’s function. Photonic cavities enhance the couplings by concentrating the electromagnetic field, which can also be understood as modifying the local density of optical states. For a single quantum emitter, this enhancement is quantified by the Purcell factor, $F_\mathrm{P} = \Gamma_\mathrm{cav}/\Gamma_\mathrm{free}$, where $\Gamma_\mathrm{free}$ is the free-space emission rate and $\Gamma_\mathrm{cav}$ is the emission rate into the cavity mode. Multiple emitters can interact through a cavity mode, leading to all-to-all connectivity.

\textbf{Many-Body Effects from QE Interactions}: Interactions between QEs can lead to a multitude of many-body effects, including quantum phase transitions, effective nonlinearities, and correlated photon emission. On-site interactions in lattices of QEs can realize Bose-Hubbard models (\textbf{g}), leading to strongly-correlated phenomena including quantum phase transitions arising from the competition between kinetic and interaction energies ~\cite{lewenstein_ultracold_2007,dutta_non-standard_2015}. Strong photon-mediated near-field dipole-dipole interactions can also lead to the Rydberg blockade effect (\textbf{h}), whereby the simultaneous excitation of multiple QEs to Rydberg states is suppressed across longer distances ~\cite{browaeys_many-body_2020}. Finally, dissipative interactions result in collective excitation structures of QE ensembles comprising both "bright" states, displaying collectively enhanced coupling to the field, and "dark" states, for which emission can be strongly suppressed or eliminated entirely (\textbf{i})~\cite{gross1982superradiance}.

\end{tcolorbox}

\begin{figure}
    \centering
    \includegraphics[width=1\linewidth]{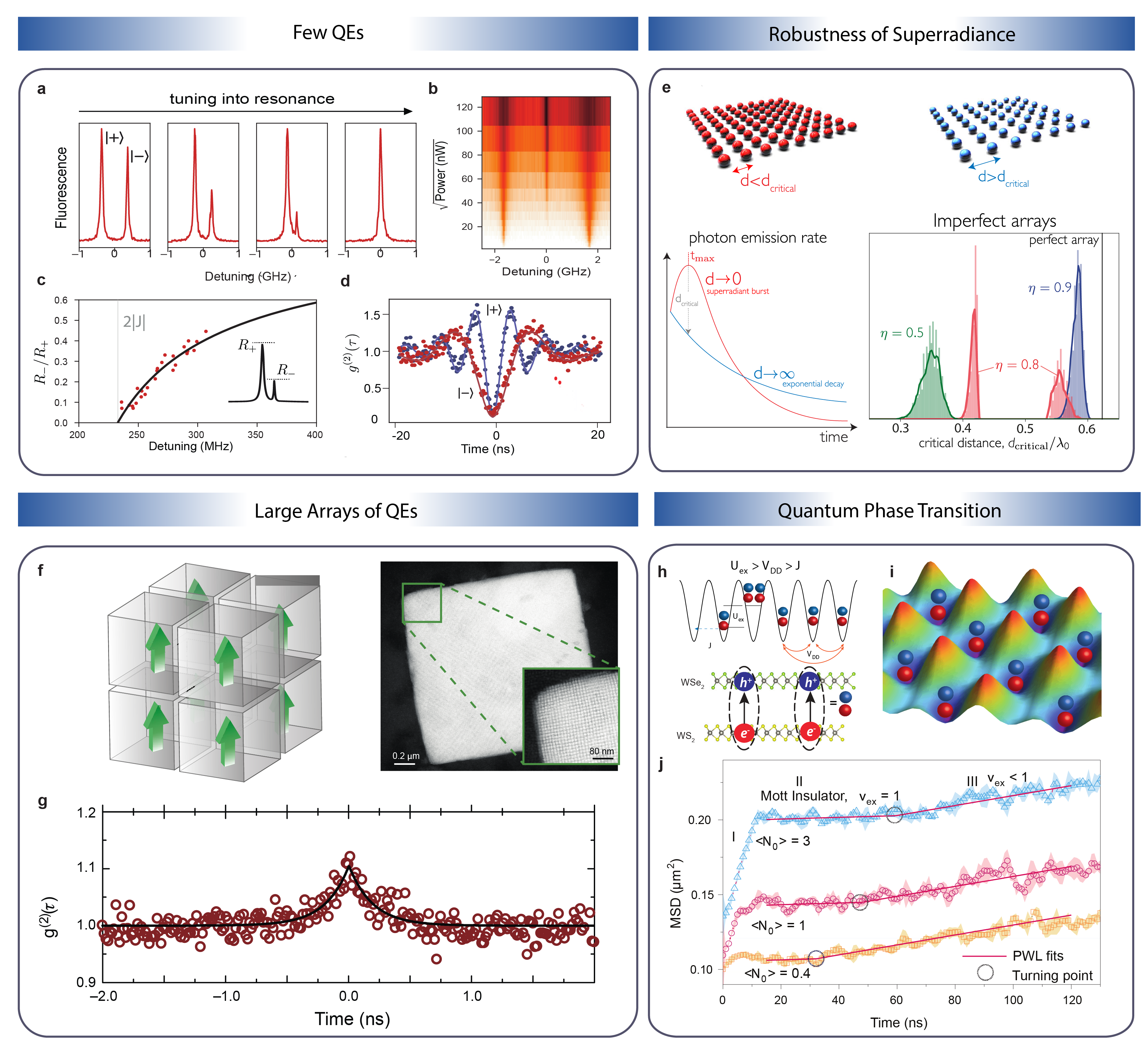}
    \caption{\emph{Many-body entanglement and collective emission in solid state QE systems.} (a) Spectra of the superradiant ($|+\rangle$) and subradiant ($|-\rangle$) single-excitation states of two sub-wavelength spaced organic molecules, showing an extinguishing of the subradiant linewidth as the molecules are tuned into resonance. (b) Fluorescence spectrum of the system in a), displaying an additional two-photon peak at sufficiently large excitation power associated with the fully-excited two-emitter state. (c) Ratio of the heights of super- and subradiant peaks in a) for varied detuning. d) Photon correlation functions for the states $|\pm\rangle$ in a), displaying modified Rabi frequencies and lifetimes. Panels a - d are adapted with permission from~\cite{lange2024superradiant}. (e) Theoretical prediction of many-body superradiance in two-dimensional sub-wavelength QE arrays. The emergence of a superradiant burst for inter-atomic spacings below a critical value is found to be robust against positional disorder (reproduced with permission from~\cite{masson2022universality}). (f) Electron micrograph of a three-dimensional perovskite quantum-dot superlattice and (g) photon correlation function $g^{(2)}(\tau)$ showing photon bunching at zero delay as a signature of superradiance in this system (panels f - g are adapted with permission from \cite{raino_superfluorescence_2018}). (h) Moiré excitons in a $WSe_2$/$WS_2$ bilayer mapped to an extended Bose–Hubbard model with strong on-site exciton–exciton interaction $U_{ex}$ and long-range dipolar coupling $V_{DD}$. Schematic potential (i) and time-resolved transport (j) reveal a crossover to a correlated (Mott-insulating) phase at high filling $V_{ex} \gtrsim 1$, evidenced by a suppressed mean-squared displacement and a turning point in the dynamics. Panels h - j are reproduced with permission from Ref.~\cite{Deng2025}.}
    \label{fig:superradiance}
\end{figure}

\subsection*{Entanglement verification and coherent control} Harnessing the entanglement of a quantum many-body state requires accurate methods of characterization. Quantum state tomography is an experimental procedure to determine the amplitudes of a quantum state, from which a desired entanglement measure can be evaluated \cite{fano1957description,wootters1989optimal}. While this provides all information of the state, the required number of samples for state tomography grows exponentially with the number of emitters, making it unsuitable for the many-body regime. Randomized measurement techniques \cite{elben2023randomized} allow the study of mixed state entanglement with much fewer samples \cite{elben2020mixed,vermersch2024many}. In fact, they can provide a classical shadow \cite{huang2020predicting} of the state: an approximate classical representation of the state from which properties such as entanglement entropies, expectation values of local observables and quantum fidelities, can be obtained efficiently. 

Under Markovian noise models, one can learn the quantum channel representing the noisy preparation using quantum process tomography \cite{chuang1997prescription,mohseni2008tomography}. This complete description of the channel, again, requires a number of samples that scales exponentially in the number of emitters. A common technique for state preparation is to start with an easy-to-prepare unentangled state and evolve it under a Hamiltonian or Lindbladian that can be written as a sum of local interaction terms. Recently, it has been shown that one can learn the Hamiltonian \cite{haah2023learning} or Lindbladian \cite{stickfranca2023efficient} coefficients with a number of samples scaling only logarithmically with the number of emitters. This means that one can efficiently check whether each local interaction term is being implemented to the desired precision.

While these methods provide very fine-grained information, their sample requirement or degree of experimental control lie beyond present capabilities of many platforms. Instead, several diagnostics of quantum behavior have been developed that are currently broadly feasible to implement. A variety of experimental techniques are used to confirm few-emitter systems are indeed forming non-classical correlations and to characterize their coherence. The most common approach is to employ photon correlation measurements, since the statistics of emitted photons, as characterized by the second-order correlation function $g^{(2)}(\tau)$, change markedly under collective emission. Specifically, an anti-dip at zero delay time which violates the upper bound $g^{(2)}(0)\leq(N-1)/N$ for emission from independent emitters provides a direct confirmation of inter-emitter correlations~\cite{auffeves_few_2011,bhatti_superbunching_2015}. This feature has already been observed in a number of systems with $N=2,3$ solid-state QEs~\cite{koong_coherence_2022,machielse_quantum_2019,grim2019scalable,kim2018super,sipahigil2016integrated,lukin2023two}. However, that recent theoretical work has shown that the observation of an anti-dip alone does not always constitute a reliable signature of superradiant behavior, since qualitatively similar features can also be observed from ensembles of non-interacting QEs under realistic experimental conditions~\cite{cygorek_signatures_2023}. Nonetheless, two-photon correlation measurements still signal non-classical light and can even quantify the potential advantage of quantum-enhanced interferometry over interferometry using a classical light source, since they allow one to efficiently compute the Quantum Fisher Information of the emitted photons \cite{abbasgholinejad2025theory}. 

Hong-Ou-Mandel (HOM) interference is the benchmark for verifying mutual coherence between distinct emitters. HOM has been used to verify coherence between remote and independent QDs: Flagg et al. tuned two separate QDs into resonance (via strain) \cite{Flagg2010} while Patel et al. achieved the same using electrical Stark tuning of remote QDs \cite{Patel2010}. By applying sequences of ultrafast pulses, one can can probe the phase coherence and manipulate the joint state of coupled emitters. Ramsey interference experiments, in which two $\pi/2$-pulses separated by a variable delay are applied, can reveal coherent oscillations between the collective eigenstates (superradiant vs. subradiant) as an oscillatory modulation in emission or absorption \cite{Bohr2024, Hotter2023}.  Such experiments are analogous to optical Ramsey control and confirm that the two-emitter system retains quantum phase coherence. 

\subsection*{Collective phenomena in large ensembles (many QEs)} Even when dealing with large numbers of solid-state emitters that are not individually addressable, striking many-body effects can emerge from their mutual interactions. In the many-body regime, highly-excited ensembles of interacting QEs are predicted to display Dicke superradiance~\cite{dicke1954coherence,gross1982superradiance}, whereby an ensemble of $N$ QEs emits a short burst of radiation with a peak intensity $\propto N^2$. There have been numerous theoretical works extending the original proposal by Dicke~\cite{dicke1954coherence}, by demonstrating that long-range photon-mediated interactions can be employed to engineer many-body superradiance in extended systems, such as sub-wavelength free-space QE arrays (see Fig.~\ref{fig:superradiance}e)~\cite{masson2022universality,masson2023dicke,sierra_dicke_2022} or QE ensembles coupled to low-dimensional fields~\cite{cardenas_many-body_2023}. Notably, it was shown that the hallmark feature of the superradiant burst is robust in such settings even in the presence of multiple competing collective decay channels and positional disorder (see Fig.~\ref{fig:superradiance}e), and that it can be predicted solely from the coherence properties of the system in the initial stages of the collective decay~\cite{masson_many-body_2020,robicheaux_theoretical_2021}. Dense solid-state QE arrays, including molecules, colloidal QDs, and 2D TMDs, are ideal platforms to experimentally test these predictions. For example, three-dimensional superlattices of nanocrystals have been shown to support superradiance emission and from many-body excitonic states (see Fig.~\ref{fig:superradiance}f-g)  ~\cite{raino_superfluorescence_2018,blach2022superradiance,Luo2025, raino_superradiant_2020,kumlin_superradiance_2025}. 

The key challenge in large ensemble of QEs regime is the characterization of the emergent entanglement structure and understanding how it can be harnessed as a resource. For example, distinguishing genuine many-body correlation from pairwise correlations remains an experimental hurdle. On the theoretical side, significant work has been dedicated to characterizing the correlation structure that emerges in the \emph{photonic} state emitted by a superradiantly decaying QE ensemble~\cite{gonzalez-tudela_deterministic_2015, gonzalez-tudela_efficient_2017}. It has been shown that superradiance generates photonic many-body states which achieve Heisenberg scaling of the quantum Fisher information~\cite{paulisch_quantum_2019,abbasgholinejad2025theory}, thereby constituting a valuable potential resource for quantum-enhanced interferometry. Suitably tailored collective light-matter coupling allows, in principle, for the realization of large decoherence-free subspaces, spanned by collective states decoupled from the field~\cite{lei2023manybodyCQED,Li2025}, providing a promising setting for the storage~\cite{asenjo2017exponential,gorshkov_universal_2007,mishra2021control} and processing~\cite{zanardi_error_1997,lidar_decoherence_1998, beige_quantum_2000,beige2000driving} of quantum information. However, it should be noted that while such subspaces are naturally protected against collective dissipation, dark states may still be subject to other decoherence mechanisms (e.g. local radiative or non-radiative decay). 

\subsection*{Quantum phase transitions} With strong interactions, a dense ensemble of QEs can enable investigations of quantum phase transitions and quantum simulation ~\cite{Kennes2021,noh_quantum_2016}. A particularly exciting platform for exploring strongly correlated many-body states of light and matter is moiré excitons in van der Waals heterostructures due to their strong exciton-exciton interactions \cite{Mak2022, Gotting2022}.  For example, when two semiconducting monolayers (such as $WSe_2$ and $WS_2$) are stacked with a slight twist or lattice mismatch, they form a long-period moiré pattern that creates an array of localized moiré excitons. The moiré potential and the exciton interaction can be tuned via twist angle and materials combination to realize different regimes of Bose-Hubbard  physics (Box 2). Individual moiré exciton traps have already shown single-photon emission, confirming that they behave as quantized emitters ~\cite{Baek2020}. Bose-Hubbard model physics has been realized using moiré excitons as well. Recent experiments have begun to reveal novel many-body phases of these excitons that arise purely from their mutual interactions~\cite{Jin2019,Marti2024, Deng2025}. Long-range dipolar repulsion freezes the motion of the Mott insulator phase for over 70 ns (see Fig.~\ref{fig:superradiance}h-j). The observed phenomenon of frozen dynamics due to strong repulsive interactions is characteristic of highly coherent systems, a feature previously realized exclusively in ultracold gases~\cite{Deng2025}. Compared with cold-atom quantum simulation platforms \cite{gross_quantum_2017}, solid-state QEs can realize stronger, longer-range interactions. For future studies, a key priority is to control many-body interactions as outlined (see Box 2) to realize diverse quantum phases, while quantifying and mitigating dephasing, such as phonon coupling, to determine how these dynamics affect phase diagrams and collective behaviors.

\section*{Photon-Mediated Many-Body Interactions}

A core requirement for entangled states of light and matter is efficient coupling between QEs and well-defined photon modes. Nanophotonic integration offers the capability to engineer light-matter interactions to achieve scalable distributed entanglement \cite{lodahl2015interfacing, lukin2020integrated, gonzalez2024light, bradac2019quantum}. The magnified coupling to a specific mode can be achieved either by increasing the spontaneous emission into the desired mode or suppressing spontaneous decay into free-space modes. If the total decay rate is increased, radiative decoherence can be accelerated to exceed dephasing from other sources, increasing an emitter's optical coherence \cite{wang2019turning}. If the Purcell-enhancing structure is coupled to a guided mode, the total collection efficiency of photons from the emitter can be increased \cite{uppu2020scalable,arcari2014near, komza2025multiplexed}. This is crucially important for generating entangled light, which is sensitive to photon loss. In the strong-coupling limit, a QE becomes a strong nonlinear element even at the level of a single photon \cite{yoshie2004vacuum, hennessy2007quantum, pscherer2021singlemolecule}. 

In the multi-emitter regime, coherent exchange of excitations via a shared cavity mode mediates effective emitter–emitter coupling, extending the interaction length scale from near-field nanometers to microns (and beyond along guided modes) as illustrated in Box 2. By optically exciting the superradiant mode of multiple QEs and projecting onto the two-spin ground states, one could prepare an entangled spin-spin state. This approach is analogous to cavity-mediated gates demonstrated in atomic systems and is a promising route toward deterministic multi-qubit logic in solid-state devices. Driving the two-photon transition of two nonidentical quantum emitters can dissipatively stabilize the system to nearly maximally-entangled stationary states \cite{vivas2023dissipative}. As nanofabrication and control improve, direct entangling gates between solid-state qubits in a photonic cavity or waveguide will become feasible. 

\subsection*{Nanophotonic resonators and waveguides} In the fast cavity regime, the enhancement of the light-matter interaction is quantified through the Purcell factor $F_\mathrm{P} = \frac{3}{4\pi^2} \left(\frac{\lambda}{n}\right)^3 \frac{Q}{V}$, where $\lambda$ is the resonance wavelength, $n$ is the refractive index, $Q$ is the cavity quality factor, and $V$ is the mode volume. The quality factor is proportional to the photon lifetime in the cavity. The mode volume is a measure of the confinement of a photon in a cavity, and a smaller $V$ results in a higher emitter-photon interaction strength. Ring resonators and disk resonators rely on highly symmetric circular geometries to achieve exceptionally high quality factors reaching into the hundreds of millions on silicon nitride platforms \cite{Ji2021ultrahigh-Q-SiN, puckett2021_422Million,Luke:13,Hosseini:09}. The tradeoff for these high Q values is a higher mode volume, which is generally tens of cubic wavelengths or larger. Photonic crystal cavities use photonic band gaps combined with total internal reflection to confine light, enabling both high Q factors from low-loss dielectric mirrors and very small mode volumes. 1D photonic crystal cavities confine light using a periodic structure along one axis (a nanobeam or nanopillar) combined with total internal reflection in the other two directions. 2D photonic crystal cavities use a 2D lattice of holes in a membrane slab to provide in-plane confinement via a photonic bandgap, while vertical confinement comes from total internal reflection. Defects in the periodicity can localize the photonic mode and concentrate the electromagnetic field within a small mode volume. Photonic crystal cavities with $Q > 100,000$ and mode volumes on the order of a cubic wavelength have been demonstrated \cite{ding2024highq}. Importantly, photonic crystal cavities such as nanobeam cavities can be strongly coupled to guided modes for efficient outcoupling of light while maintaining high Q factors. 

Light-matter interactions can be enhanced by slowing the group velocity using photonic crystal waveguides (PCW), with demonstrated slow-down factors up to 300 \cite{vlasov2005active}. For a PCW constructed with a line defect in a 2D photonic crystal, an emitter experiences strong suppression of free-space emission, leading to high collection efficiency even with modest increases in emission rate. PCWs also offer broadband operation, providing versatility across different wavelength ranges. Furthermore, PCWs can be engineered to support chiral light-matter interactions where light propagation becomes unidirectional due to optical spin-orbit coupling \cite{lodahl2017chiral, sollner2015deterministic}. Such chiral coupling enables applications in non-reciprocal photonic devices (e.g., optical isolators and circulators) and establishes directional coupling between multiple emitters embedded in the same waveguide, forming cascaded quantum systems.

Plasmonic structures exploit collective oscillations of surface electrons in metallic nanostructures to confine electromagnetic energy below the diffraction limit and produce strong localized field enhancements \cite{Gramotnev2010-ye}. Purcell factors upward of 10,000 are achievable in structures such as nanoparticle-on-mirror cavities or nanoparticle antennas \cite{Akselrod2014-eo, Bogdanov2018mm, ng2019plasmonic}. A major limitation of localized plasmonic systems is that much of the Purcell enhancement is non-radiative: energy stored in plasmonic modes is often converted into heat through Ohmic losses instead of being emitted as photons. One strategy to mitigate this effect is the use of hybrid dielectric–plasmonic cavities, which combine the high Purcell factors of plasmonic resonances with low-loss dielectric structures, enabling efficient photon outcoupling before energy is dissipated as heat \cite{andersen2018hybrid}. Another approach involves plasmonic nanoparticle lattices, where localized surface plasmon resonances are coupled to the diffractive orders. In addition, careful material choices, such as low-loss metals~\cite{Kocoj2025-ln,Gao2023-vx}, can further mitigate material instability and thermal dissipation.

Modified dielectric cavities can also achieve extreme confinement, reaching ultra-small mode volumes as low as $7\times10^{-5} \, (\lambda/n)^3$ \cite{choi2017selfsimilar}. Another growing class of photonic structures are generated through global topological optimization, rather than the adiabatic modification of photonic crystals to create defect states. Topology-optimized (or inversely-designed) cavities have demonstrated similar mode volumes of $3\times 10^{-4} \, (\lambda/n)^3$ with small device footprints \cite{albrechtsen2022nanometerscale}. 

Across all platforms, materials and integration are of critical importance. High index contrast materials tighten confinement but may increase loss; embedding the emitter during growth maximizes light-matter coupling but restricts the choice and density of QEs, whereas hybrid integration eases assembly at some cost to positioning accuracy and coupling strength. The optimal choice depends on target applications, uniform coupling across many emitters, bandwidth versus coherence, and the required degree of programmability for many-body interactions.

\begin{figure}
    \centering
    \includegraphics[width=1\linewidth]{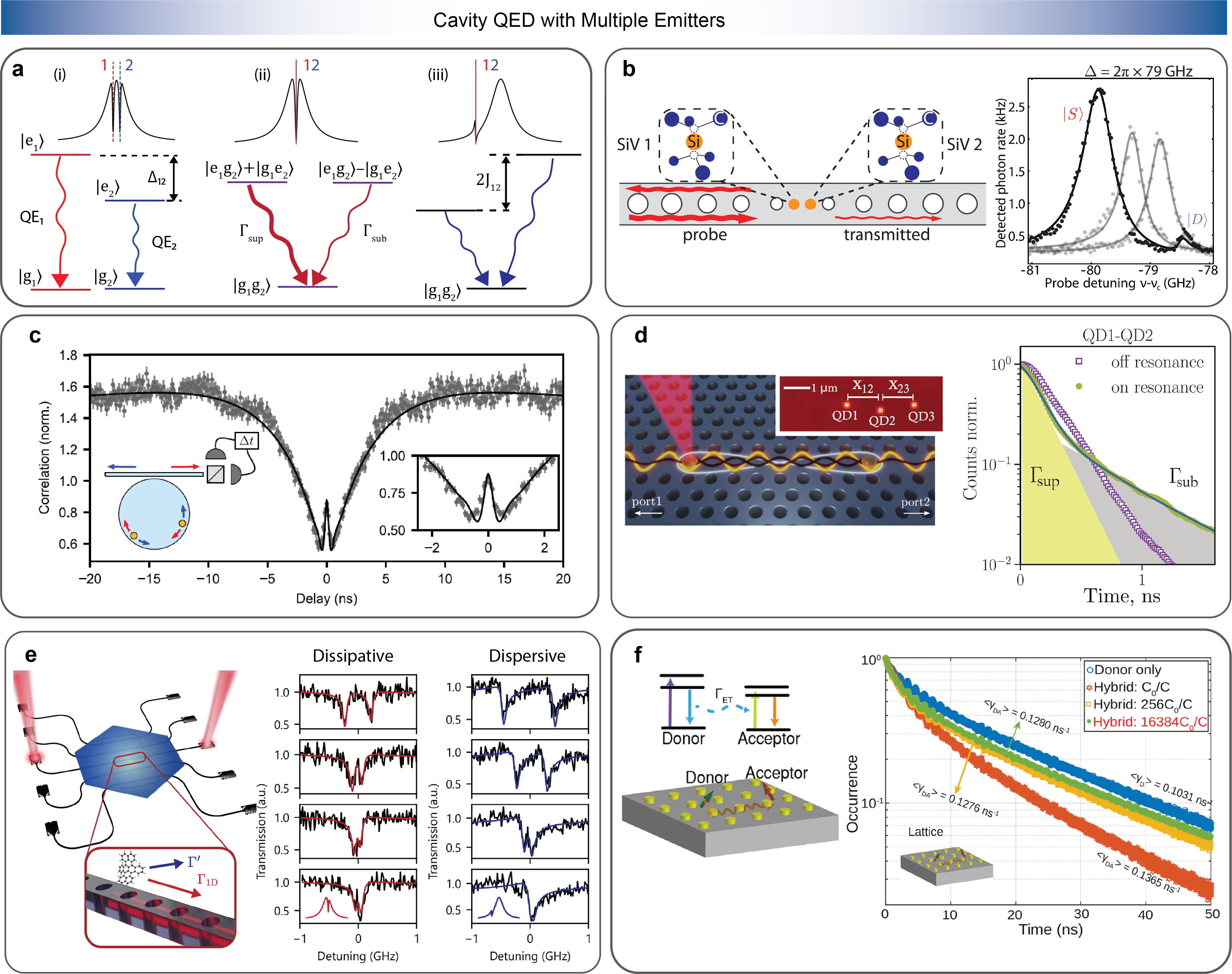}
    \caption{\emph{Collective interactions of multiple solid-state quantum emitters in nanophotonic environments observed across molecular, quantum-dot, and color-center platforms.} (a) Schematic cavity transmission spectra and energy-level diagrams illustrating three regimes: (i) two emitters off resonance with each other but near-resonant with the cavity, decaying independently, (ii) emitters resonant with each other and with the cavity (dissipative regime), where superradiant and subradiant collective states form, and (iii) emitters resonant with each other but detuned from the cavity (dispersive regime), where coherent dipole–dipole interactions split the symmetric and antisymmetric states. (b) Cavity-mediated interaction of two silicon-vacancy (SiV$\hspace{0pt}^-$) centers in a diamond nanocavity, showing formation of bright (superradiant, |S⟩) and dark (subradiant, |D⟩) collective states, reproduced with permission from Ref. \cite{evans2018photon}. (c) Superradiant emission observed from two V$_{si}$ centers in a thin-film SiC microdisk cavity manifested through bunching in photon correlation measurements, reproduced with permission from Ref. \cite{lukin2023two}. (d) Observation of super- and subradiant emission from pairs of quantum dots deterministically positioned in a nanophotonic waveguide, mediated by long-range radiative coupling, reproduced with permission from Ref. \cite{tiranov2023collective}. (e) Two dibenzoterrylene (DBT) molecules coupled to a 1D photonic crystal cavity, interacting either dissipatively via the cavity mode or dispersively through dipole–dipole coupling, reproduced with permission from Ref. \cite{lange2025cavity}. (f) Long-range dipole-dipole interactions mediated by plasmonic nanoparticle lattice manifested through decay lifetime at various acceptor concentrations, reproduced with permission from Ref. \cite{boddeti2022dipole-dipole}.}
    \label{fig:placeholder}
\end{figure}

\subsection*{Cavity QED with multiple emitters} Cavity QED with multiple QEs coupled to a common photonic mode unlocks collective phenomena central to many-body quantum optics. A remarkable feature of cavity and waveguide QED is that QEs can interact coherently without requiring nanometer-scale separations (Box 2). The photon-mediated interactions, described by the Tavis-Cummings model, give rise to collective eigenstates and modified decay rates. These effects were first observed in atomic ensembles and have now been realized in solid-state platforms as well ~\cite{asenjo2017exponential,Sheremet2023,Guerin2016,Das2020,otten2015entanglement,Asaoka2022,McDonnell2022}. 

Two-emitter cavity QED can be understood in three regimes (Fig.3 a): (i) when the emitters are detuned from each other but near-resonant with the cavity, they act independently and each experiences the ordinary Purcell enhancement; (ii) when the emitters are resonant with each other and the cavity, they form symmetric (superradiant) and antisymmetric (subradiant) collective states, with the superradiant state decaying at an enhanced Purcell rate while the subradiant state decouples from the cavity mode; and (iii) when the emitters are resonant with each other but detuned from the cavity, they couple dispersively via a virtual photon exchange, leading to a dipole–dipole interaction that shifts the symmetric and antisymmetric states by the exchange rate  $\pm J_{12}$, producing a splitting of $2J_{12}$.

Solid-state demonstrations of radiative emitter–emitter coupling in cavity QED were achieved first using SiV centers in diamond.  Evans et al. demonstrated two SiV centers embedded in the same nanophotonic diamond cavity (Fig. 3b) \cite{evans2018photon}. Each SiV was individually strongly coupled to the high-Q cavity mode. The common cavity mode induced a coherent interaction between them, manifesting as a pronounced splitting in the optical spectrum due to coherent dipole–dipole coupling mediated by cavity photons. By using each SiV’s electronic spin as a controllable qubit, the interaction can be turned on and off (by tuning one emitter in and out of resonance). Lukin et al. reported the integration of V$_{si}$ centers into thin-film silicon carbide microdisk resonators fabricated on a SiC-on-insulator platform \cite{lukin2023two}. The high-Q whispering-gallery modes of the microdisks provided strong Purcell enhancement, yielding single-emitter cooperativities up to 0.8. When two V$_{si}$ centers coupled to the same resonator mode, their joint emission exhibited superradiant bunching in photon correlation measurements, confirming collective radiative decay (Fig. 3c).

Cavity QED with two epitaxial QDs has also been demonstrated. A foundational step toward multi-emitter applications was demonstrated by observing coherent coupling and collective super- and subradiant emission from deterministically positioned QDs in a nanophotonic waveguide (Fig. 3d) \cite{tiranov2023collective}. These experiments showed clear signatures of photon-mediated interactions, including enhanced collective emission rates (superradiance). By tuning individual QD emission wavelengths via electrostatic gating, magnetic field or strain engineering, it becomes possible to map out interaction-driven transitions between collective eigenstates.  Steady-state subradiant emissions from two InAs QDs embedded in a low-Q, highly directional circular grating Bragg cavity has also been achieved, where the subradiant-state population dominates the steady-state with a highly negative cooperativity ~\cite{Kim2025}. Long-range energy transfer between two InAs QDs embedded in a PCW was demonstrated by tuning the emitters into resonance and measuring intensity autocorrelations with $g^{(2)}(0)$= 0.94 ~\cite{Diepen2025}. 

In addition to color centers and QDs, recent work has demonstrated the compatibility of organic molecules with nanophotonic resonators. Rattenbacher et al. coupled lifetime-limited dibenzoterrylene molecules in polyethylene films to disc resonators, using electrodes to tune molecular pairs into resonance \cite{rattenbacher2023onchip}. Lange et al. coupled dibenzoterrylene molecules in anthracene crystals to nanobeam cavities, employing permanent optically-induced Stark shifts to achieve resonant coupling in both dissipative and dispersive regimes (Fig. 3e) \cite{lange2025cavity}. The high density achievable with molecular emitters can, in principle, enable coherent coupling among large ensembles of emitters (> 3).

Finally, plasmonic cavities can also mediate many-body entanglement. An isolated nanoparticle antenna can tightly confine the light field and support pairwise or small-scale entanglement among a few emitters, though the efficiency depends heavily on emitter number and position due to rapid near-field decay \cite{PhysRevLett.121.173901,PhysRevA.86.011801,PhysRevB.89.235413}. The uniformity of the field distribution can be improved through nanoparticle-on-mirror cavities, that exhibit nanoscale dielectric gaps that produce spatially homogeneous fields~\cite{Leng2018-sm,PhysRevB.101.035403}. Such designs enable multiple quantum emitters to experience comparable coupling conditions, and consequently enhanced overall coherence \cite{PhysRevA.111.013526,PhysRevB.111.075420}. Plasmonic nanoparticle lattices support delocalized surface lattice resonances (SLRs) to facilitate coherent interactions across many unit cells and over hundreds of nanometers. For instance, Boddeti et al. experimentally demonstrated this principle using silver nanoparticle arrays coupled to donor and acceptor dye molecules (Fig. 3f) \cite{boddeti2022dipole-dipole}. Angle-resolved spectroscopy and fluorescence lifetime measurements revealed resonance energy transfer that persists up to about 800 nm of mean donor–acceptor separation, which is nearly two orders of magnitude larger than free space. Such long-range dipole–dipole coupling mediated by SLRs open up opportunities for exploring collective phenomena such as superradiance and correlated photon emissions \cite{boddeti2024reducing}. Moreover, when the lattices are strongly coupled with emitters, the formed exciton-polaritons inherit the nonlinearity of excitons and can achieve polariton lasing and condensation at room temperature~\cite{Freire-Fernandez2024-vv,Hong2025-gb,De_Giorgi2018-fu,Ramezani:17}.

Collectively, these results shown in Fig. 3 demonstrate that with nanophotonic integration, even solid-state emitters separated by many wavelengths can interact strongly \cite{papon2023independent,chu2023independent}. In future studies, pushing cavity QED from pairs to many solid-state QEs hinges on making the collective Hamiltonian programmable and homogeneous while keeping each emitter transform-limited. This requires further material developments in the deterministic placement and active frequency control of emitters as well as strategies and designs of nanophotonics that improve light-matter coupling and collection efficiencies.    

\section*{Photon Nonlinearity}
Photons can not only serve as a mediator of entanglement between emitters but can also carry quantum entanglement as resource states themselves. Therefore, strong photon–photon interactions at the single- or few-photon level represent a major milestone in quantum optics. Achieving this enables all-optical analogues of transistors and logic gates at the quantum level, as well as deterministic entangling operations between photons \cite{chang2014quantum,Imamoglu1997}. Photons do not naturally interact with each other in free space or linear media, so achieving photon nonlinearities requires mediating interactions through matter. Early theoretical proposals showed that embedding resonant emitters or nonlinear media in optical systems could induce an effective photon–photon interaction \cite{lukin2001dipole,chang2014quantum}. The main underlying mechanism is an excitation blockade that can originate either from the saturation of a single or a few two-level emitters, or from strong and long-ranged interactions, as can be mediated by Rydberg states (Rydberg blockade, as illustrated in Box 2).   
In the strong light-matter coupling limit or in combination with electromagnetically induced transparency (EIT), the nonlinearity of the emitters can be translated into a photon-photon interaction, which can even yield the limit of a photon blockade wherein one photon in a nonlinear medium prevents a second photon from entering 
~\cite{Imamoglu1997,Miranowicz2013}. These ideas laid the groundwork for quantum nonlinear optics, suggesting that a single photon could switch or modulate another. 

Large photon nonlinearities require both a strong light–matter interaction to imprint the presence of each photon onto a matter excitation and a strong matter–matter interaction so that one excitation significantly shifts or blocks a second. By simultaneously engineering high‑cooperativity photon coupling (via microcavities, waveguides, or plasmonic confinement as discussed in the previous section) and enhancing emitter–emitter interactions (through Rydberg excitation, moiré trapping, or dipole‑dipole coupling as discussed in Box 2), recent experiments have pushed nonlinear responses into regimes previously accessible only with intense light fields \cite{kala2025opportunities}.

\begin{figure}
        \centering
        \includegraphics[width=1\linewidth]{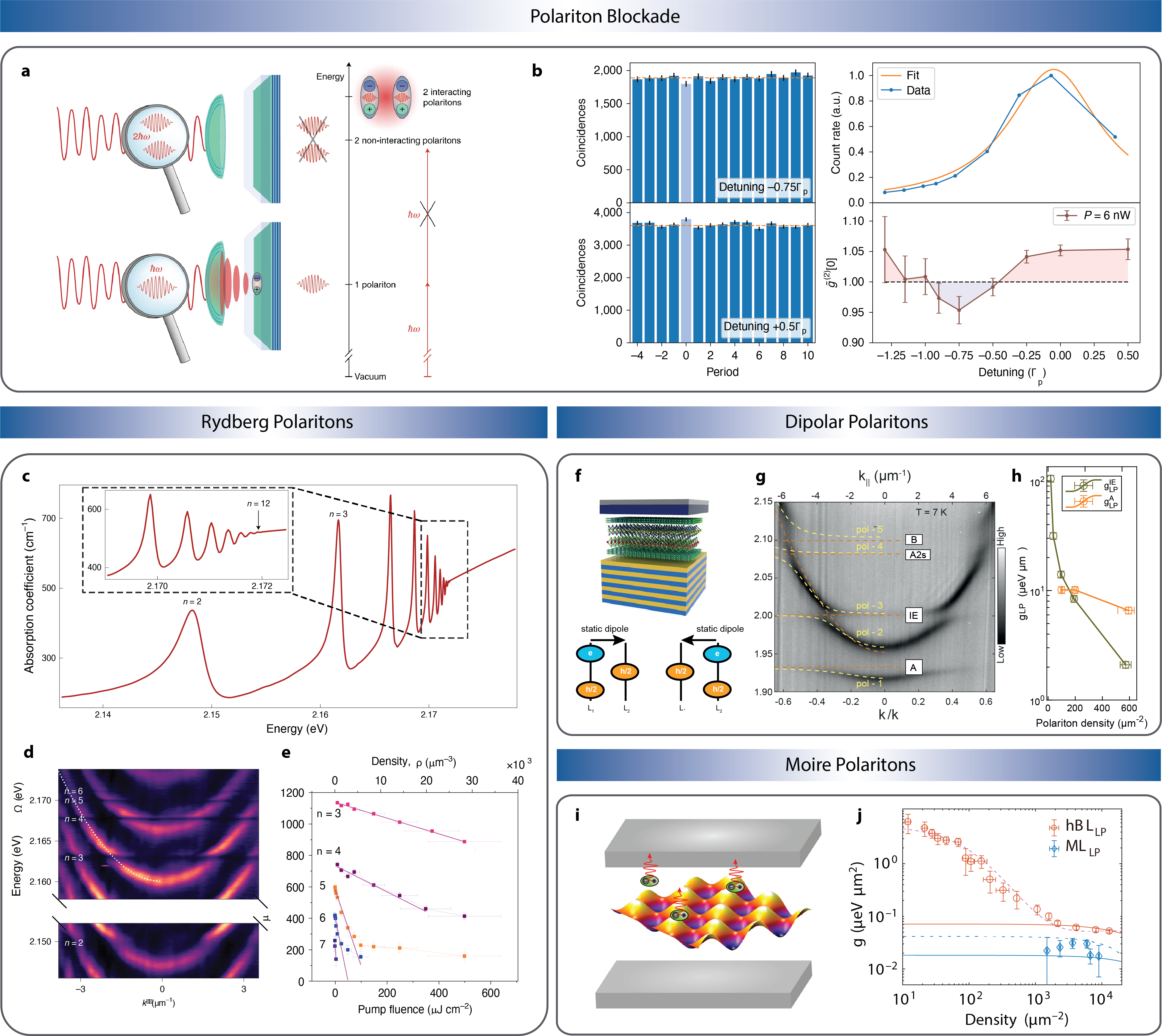}
        \label{Figure 4}
  
    \caption{\emph{Pathways to achieve strong photon nonlinearity through exciton polaritons.} (a) Excitation and re-emission of an exciton polariton by a single photon.  Interactions shift the mode off resonance, suppressing multi-photon transmission (polariton blockade), reproduced with permission from Ref. \cite{gerace2019quantum} (b) Detuning dependent coincidence counts and the corresponding nonclassical correlations from polaritons based on quantum well microcavity, reproduced with permission from Ref \cite{delteil2019}. (c) (Top) Absorption in a natural Cu$_2$O crystal resolving excitonic resonances for large principal quantum numbers. (d) Formation of Rydberg exciton polaritons under strong coupling regime for n=3,..,6. Panels c - d are reproduced with permission from Ref. \cite{orfanakis2022} (e) Rabi splitting vs. exciton density and pump fluence is shown for different principal quantum numbers $n$, revealing an increase in $n$-dependent nonlinear coefficient, reproduced with permission from Ref. \cite{Makhonin2024}. (f) A bi-layer MoS2 coupled to Fabry-Perot microcavity. (Bottom Left) Interlayer excitons with permanent out-of-plane dipole and strong, long-range interactions. (g) Polariton dispersion showing multiple coupled excitonic species. (h) Interlayer (dipolar) polaritons exhibiting enhanced nonlinearity compared to neutral exciton polaritons, evident in low density reduction of apparent Rabi splitting (panels f - h adapted with permission from Ref. \cite{Datta2022}). (i) Moiré polariton system formed by exciton confined in a moiré lattice, coupled to cavity. (j) Nonlinear enhancement of moiré excitons over intralayer exciton polaritons, (panels i - j are reproduced with permission from Ref. \cite{Zhang2021}). }
\end{figure}

\subsection*{Polaritons with Rydberg excitons} One of the most advanced platforms for single-photon-level nonlinearity is based on ultracold atomic ensembles with Rydberg excitations ~\cite{Gorshkov2011,chang2014quantum}. In this approach, photons are coupled via EIT to highly excited Rydberg states of atoms, forming Rydberg polaritons. When one photon excites a Rydberg atom, it can block the excitation of nearby Rydberg atoms within a “blockade radius,” effectively causing two photons within that range to interact strongly. This mapping of Rydberg-atom interactions onto photons has enabled a host of quantum nonlinear optical effects. Notably, it allows single photons to phase-shift or attenuate other photons ~\cite{Kim2021,Pritchard2010} and even implement two-photon gates \cite{tiarks2019, stolz2022}. The trade-off, however, is that such experiments typically require ultracold temperatures or vacuum systems and precise control of atomic states ~\cite{Peyronel2012,Dudin2012,Firstenberg2013}.

In the solid-state realm, an exciting development is the use of Rydberg exciton-polaritons in semiconducting materials to achieve photon nonlinearity (Fig. 4a-b). Exciton-polaritons arise in semiconductors or two-dimensional materials when an exciton strongly couples to a cavity photon, forming a hybrid quasiparticle. Polaritons can interact with each other via their excitonic component, and if these interactions are strong enough, quantum nonlinear effects analogous to those in atomic systems can emerge. Fig. 4b shows a small experimental dip in $g^{(2)}(\tau)$ at $\tau$=0 ($g^{(2)}(0)<1$), demonstrating the onset of a polaritonic excitation suppression in a semiconductor quantum-well microcavity~\cite{delteil2019}. 
Rydberg excitons in bulk semiconductors have been pursued as an analog to atomic Rydberg gases, where the excitation suppression can potentially be dramatically enhanced~\cite{walther2018giant}. Cuprous oxide (Cu$_2$O) has emerged as one of the most successful material candidates: it supports Rydberg excitons with principal quantum numbers up to n=30, whose diameters can approach 1 $\mu$m (Figure 4c). These Rydberg excitons in Cu$_2$O have extremely large electric transition dipole moments, leading to exciton–exciton interaction strengths orders of magnitude larger than among ground-state excitons. Giant Rydberg excitons in Cu$_2$O have been experimentally demonstrated in bulk crystals~\cite{Kazimierczuk2014,Lynch2021,Paul2024}, synthetic thin films~\cite{Steinhauer2020,DeLange2023}, and two-dimensional arrays~\cite{Barua2025}. Their power and temperature-dependent characteristics, as well as their controllability through microwave fields \cite{Gallagher2022} have been investigated \cite{heckoetter2021, walther2020, Morin2022}, providing important insights into the linear and nonlinear optical properties of this system. Orfanakis \textit{et al.} \cite{orfanakis2022} achieved strong light coupling to Cu$_2$O Rydberg excitons up to n=6, creating the first Rydberg exciton-polaritons (Fig. 4d). Kerr-type optical nonlinearities arising from the Rydberg blockade effect in a Cu$_2$O microcavity have been demonstrated ~\cite{Makhonin2024}. The polariton resonance frequency was found to shift depending on the polariton density (Fig. 4e), an effect attributed to Rydberg exciton blockade (one exciton preventing nearby excitations). 

Monolayered TMDs host tightly bound excitons with large binding energies, and they can sustain polaritons even at room temperature \cite{liu2015strong,sun2017optical}. Though larger binding energies are typically associated with smaller interaction strengths, the two-dimensional material geometry offers unique opportunities for the realization of photon-photon interactions. Combination with a microcavity~\cite{Sina2018} or a two-color excitation scheme \cite{walther_nonclassical2022} has been proposed as a way to exploit the material structure to produce light with quantum mechanical statistics. Gu \textit{et al.} experimentally realized the formation of polaritons based on the 2s exciton in monolayer $WSe_2$, and compared their nonlinear properties with those of polaritons formed from the ground 1s exciton ~\cite{Gu2021}. Because excited-state excitons are larger in size, their mutual interactions are stronger. However, the enhanced nonlinearity of the $2$s-state still falls short of the “quantum blockade” regime. Rapid dephasing and strong phonon scattering in TMDs pose significant challenges in accessing higher Rydberg states, representing a major obstacle to achieving single-photon nonlinearity.

\subsection*{Dipolar and moiré exciton polaritons } Another promising strategy to strengthen exciton–exciton interactions is to use excitons with a permanent electric dipole moment, enabling dipolar interactions. One such example are interlayer excitons, which can form in van der Waals bilayers of TMDs due to the separation of an electron-hole pair across the layer interface. Large nonlinear interactions between interlayer exciton polaritons have been demonstrated with bilayer MoS$_2$~\cite{Datta2022} (Figure 4f-h). As described in the Many-Body Interactions, Collective Photon Emission and Quantum Phase Transition section, moiré excitons feature reduced bandwidth and very large nonlinearity due to spatial localization, and strong onsite interactions in addition to long-range dipolar interaction. Moiré exciton-polaritons also exhibit enhanced and electrically tunable nonlinearities~\cite{Zhang2021} (Figure 4i-h). The moiré approach offers a tantalizing route toward scalable quantum nonlinear optics on a chip with an array of interacting polariton sites that could mimic a strongly correlated photonic lattice, all in a two-dimensional solid.

\subsection*{Strategies to further enhance interactions} Taking inspiration from ultracold atomic gases, researchers have sought Feshbach resonances in excitonic systems, where a two-particle scattering interaction is dramatically enhanced by the presence of a bound state at nearly the same energy ~\cite{Takemura2014}. A bound state of two excitons is a biexciton and a bound state of an exciton and an extra charge is known as a trion. In GaAs quantum wells, it was demonstrated that when two polaritons with opposite spin were tuned into resonance with the biexciton energy by adjusting the cavity detuning, their effective interaction switched from weakly repulsive to strongly attractive and then back to repulsive as the resonance was crossed~\cite{Takemura2014}. In TMD monolayers, biexcitons exist as well (with binding energies on the order of 20–40 meV), so in principle, a similar tuning could be achieved by bringing the polariton or two-exciton energy near the biexciton state. Due to the larger binding energy, the resonance condition in TMDs might be achieved at higher temperatures or with smaller detunings. Moreover, recent theoretical proposals suggest using trion states in a controlled way to induce Feshbach-like tunability~\cite{Ido2021,Wagner2025}. To enhance light-matter interaction, using photonic structure to further confine the photons or slow down photon propagation would be desirable. TMDs could offer a simpler route to implement such structures: their atomically thin nature means they can be transferred onto virtually any photonic platform without perturbing the mode structure. This 2D flexibility has enabled TMD strong coupling in a wide variety of architectures as discussed in Photon-Mediated Many-Body Interactions section, from planar cavities and waveguides to nanogaps and plasmonic antennas~\cite{Schneider2018,Luo2024}.

\section*{Generation and Characterization of Cluster and Graph states}
Graph states are special types of entangled quantum states that can be visualized as graphs, with the vertices representing the qubits and the edges between connected qubits representing entanglement links (see e.g. Fig.~1, and Ref.~\cite{hein2006entanglement} for a review). In particular, the collection of nodes and edges mathematically describes a complete set of correlation observables that uniquely determine the corresponding graph state. The most well known graph state is the so-called cluster state, a 2D square graph, which was shown to be a universal resource for quantum computing in the seminal paper by Raussendorf and Briegel~\cite{raussendorf2001one}. Photonic graph and cluster states are of special interest, as their use can overcome challenges associated with photonic quantum computing and quantum repeaters. Generating large photonic cluster or graph states has been difficult to scale. The conventional approach utilizes probabilistic linear-optical methods; independent single photons or entangled photon pairs, typically generated through non-linear processes such as spontaneous parametric down-conversion (SPDC) or spontaneous four-wave mixing (SFWM), are interfered, and a subset of them measured, to “stitch” photons into a cluster state \cite{browne2005resource}. Standard objections to the probabilistic approach have been: ($i$) The intrinsically low efficiency of SPDC and SFWM entangled photon pair sources, and ($ii$) the probabilistic nature of the approach (the protocol succeeds conditional on certain low-probability measurement outcomes); related to this, the fusion gates for photons are themselves probabilistic \cite{browne2005resource}. These challenges are in part addressed by ($i$) pumping the SPDC sources with an ultra-fast pump laser, which increases the entangled pair generation rate \cite{zhong201812}, as well as ($ii$) reusing the contracted graph states following a failed fusion gate allowing the gate to be re-attempted, which reduces the resource requirements (in terms of entangled photon pairs) for a successful fusion event. Although this enabled pioneering demonstrations of photon cluster states \cite{walther2005experimental}, this approach remains difficult to scale \cite{varnava2008good, pant2017rate} without combining it with deterministic methods, as done in recent experiments \cite{thomas2022efficient, thomas2024fusion}. Recent research has therefore turned to a combination of probabilistic and deterministic protocols~\cite{hilaire2023graph} or fully deterministic sources: solid-state QEs with spin–photon interfaces that can emit or mediate entanglement on demand \cite{lindner2009proposal, lee2019quantum, economou2010optically,gimenosegovia2019}. In principle, this could be done with single or multiple QEs. Below, we review recent advances in generating photonic cluster/graph states deterministically using solid-state QEs, focusing on epitaxial QDs \cite{cogan2023deterministic, coste2023high} and color-center spins in diamond and silicon, and we discuss how these states can be verified and characterized.

\subsection*{Generation of cluster states at spin-photon interface} Experiments with atomic emitters improved the quality of the gates and demonstrated a record number of photons in a linear cluster state \cite{schwartz2016deterministic} and a Greenberger-Horne-Zeilinger (GHZ) state.  However, the naturally longer timescales and technically challenging atomic systems still limited the generation rate and scalability. Later experiments with QDs improved on this (Fig. 5a-h) \cite{cogan2023deterministic}, including recent work with a cavity enhancing the emission (Fig. 5a-d) \cite{coste2023high, huet2025deterministic}. It is therefore evident that solid-state QEs can generate photonic graph states fast while providing the potential for device integration.

Aside from epitaxial QDs, color centers are particularly well suited for generating graph states beyond a 1D cluster state due to the presence of nearby nuclear spins with optically accessible initialization, control, and readout through the defect spin. In addition, the ability to perform conditional gates between the spin defect and nearby nuclear spins has been demonstrated both for strongly and weakly coupled nuclear spins. The nearby strongly coupled nuclei can be addressed directly by using the Overhauser shift of the defect energy levels, resulting in relatively fast entangling gate times. In contrast, entangling gates with weakly coupled nuclei \cite{taminiau2012detection, van2012decoherence} require applying a sequence of pulses with an inter-pulse delay chosen to allow selective coupling with nuclei of particular hyperfine coupling strengths. This leads to relatively long entangling gate times, but allows for the possibility of multi-qubit entangling gates~\cite{PhysRevX.13.011004, takou2}, which can perform better, as also shown experimentally \cite{Minnella2025}, than sequential nuclear spin entangling gates~\cite{cramer2016repeated}. The latter approach is not well-suited for a dense nuclear spin bath, such as in QDs, and performs better with sparse baths where the number of nuclear spins of a particular hyperfine strength is relatively low, reducing the number of unwanted interactions. Such baths are characteristic of diamond and silicon, which have a natural abundance of nuclear spin isotopes $ ^{13}$ C and $ ^ {29}$ Si, respectively, and the different nuclear spin species can be beneficial both in terms of coherence  times \cite{Widmann2014,Bourassa2020} and gate fidelities~\cite{dakis2024}.  Conditional gates between the defect electronic spin and nearby as well as far nuclear spin gates have been demonstrated experimentally for NV centers. However, due to their small Debye-Waller factor of $\approx 3\%$, they are not best suited for operation as quantum emitters.  In contrast, recent interest has been increasing for defects in Silicon, such as T and G centers \cite{bergeron2020silicon,durand2021broad,udvarhelyi2021identification}, which emit in the telecommunication band, making them optimal for fiber-optical communication. These emitters also exhibit small Debye-Waller factors (approximately $23\%$ and $ 18\%$, respectively), but recent experiments have demonstrated improvements when embedded in cavities, e.g., for T-centers \cite{islam2023cavity,johnston2024cavity} and G-centers \cite{redjem_all-silicon_2023,saggio2024cavity}. By utilizing spin defects as quantum emitters and the surrounding nuclear spins as ancilla qubits, the latter can temporarily store the entanglement with already emitted photons, while the former generates new parts of the graph state \cite{buterakos2017deterministic}. For example, it has been shown that the presence of a nuclear spin near a solid-state defect emitter suffices to generate resource states for quantum networks \cite{buterakos2017deterministic} and a protocol for the generation of arbitrary photonic graph states has also been developed~\cite{li2022photonic}.

\subsection*{Characterization of entanglement structure} Verifying that one has indeed generated the desired cluster or graph state – and characterizing its entanglement structure – is a non-trivial task that generally grows harder with the number of photons. The “gold standard” method is quantum state tomography, wherein a large set of measurements in different bases is collected to reconstruct the full multi-qubit density matrix. Tomography provides complete information, allowing one to compute state fidelity, entanglement measures, and observe the exact structure of entanglement (e.g. which qubits are pairwise entangled, etc.). Its large measurement overhead makes it impractical in the many-body setting. Instead, specialized measures that leverage the stabilizer structure of graph states can be estimated more efficiently.

Two common strategies to verify entanglement in a state are estimating the fidelity between the experimental state and the desired graph state and estimating an entanglement witness. Lower bounds in the fidelity can be estimated efficiently for GHZ states, graph and cluster states by local measurements of a small set of stabilizer generators \cite{guhne2007toolbox,toth2005entanglement,guhne2009entanglement,knips2016multipartite}. Under realistic noise models, the lower bound on the fidelity can be improved significantly while requiring a number of samples only linear in the number of qubits \cite{tiurev2022highfidelity}. The fidelity with an entangled stabilizer state is intimately linked to the verification of entanglement in the state using an entanglement witness $W$: this is an operator whose expectation value satisfies $\text{Tr}(W\rho) \geq 0$ for all states that are separable across some bipartition. Thus, if one finds $\text{Tr}(W\rho)<0$, the state $\rho$ exhibits genuine multipartite entanglement. The expectation value can be estimated with the same stabilizer measurements used for the fidelity lower bound, so no extra overhead is incurred. This kind of entanglement witnesses have been used to certify the multipartite entanglement in 14-photon GHZ states and 12-photon linear cluster states \cite{thomas2022efficient}. 

While these strategies provide a lower bound on the fidelity, this bound is not necessarily tight. Instead, one might want to estimate the infidelity to a desired precision with high confidence. This can be estimated by measuring random sequences of local Pauli measurements \cite{pallister2018optimal,dangniam2020optimal}. The number of sequences is independent on the size of the system and depends mildly on the precision and confidence.

\begin{figure}
    \centering
    \includegraphics[width=1\linewidth]{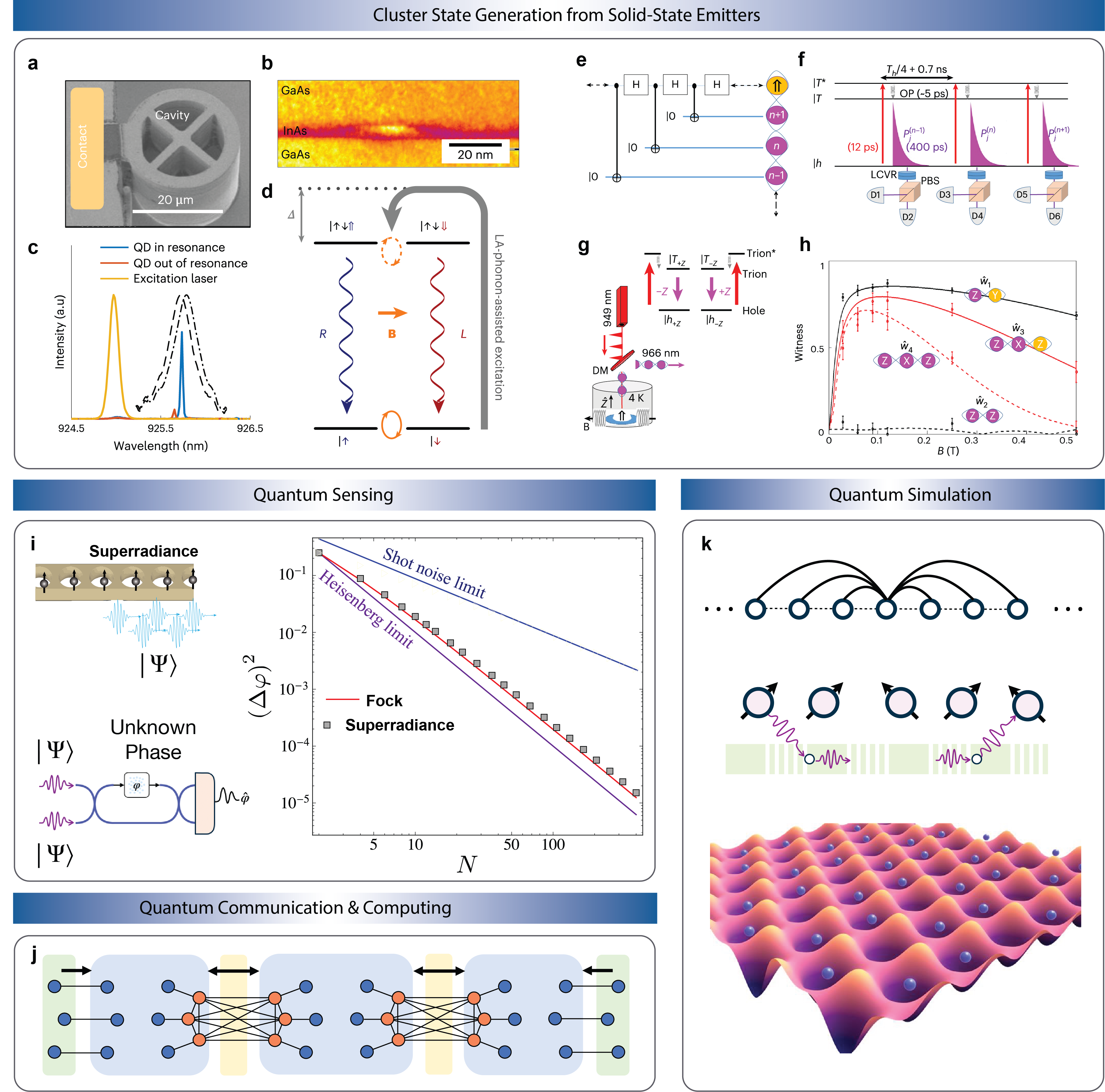}
    \caption{\emph{State of the art generation of Graph States and applications of many-body entangled states.} (a-h) Generation of cluster and graph states from QDs. (a) Negatively charged QD in the center of a connected pillar optical cavity, (b) single InGaAs QD with GaAs barriers, (c) QD spectral emission, (d) Energy levels and
    optical selection rules of the negatively charged QD in the presence of a small
    (<100 mT) transverse magnetic field. Panels a - d are reproduced with permission from Ref. \cite{coste2023high}. (e) cluster state generation using sequential Hadamard and CNOT gates, (f) QD spin configurations, optical transitions, and their detection used in the measurements, (g) DM represents a dichroic mirror and the magnetic field is applied along the $x$-direction, the states refer to the hole, trion, and excited trion states, (h) the measured (error bars) and calculated (coloured lines) cluster state witnesses. Panels e - h are reproduced with permission from Ref. \cite{cogan2023deterministic}. (i) (Left) The application of many-body entangled states for quantum-enhanced sensing of an unknown phase (see \cite{pezze2018quantum}). (Right) Superradiant emission allows for detection that surpasses the shot noise level, image adapted with permission from Ref. \cite{paulisch_quantum_2019}. (j) Graph states as resource states for quantum communication and quantum computation, generated by specialized quantum emitters, such as QDs, adapted with permission from Ref. \cite{zhan2023performance}. (k) The presence of long-range interactions between qubits allows for performing quantum simulation of complex many-body systems.}  
    
    \label{fig:placeholder}
\end{figure}

\section*{Outlook/Future Perspectives}

Engineering many-body entanglement in emitter arrays and the emitted photons will serve as a fundamental resource for all three main quantum technologies: quantum computation, communication, and sensing (Fig. 5i-k).  In sensing applications (Fig. 5i), twin Fock states \cite{holland1993interferometric, sanders1995optimal} and NOON states \cite{lee2002quantum, dowling2008quantum} have long been known to be optimal for phase estimation, as measured by the Heisenberg scaling of the estimation mean-squared error (NOON states are also commonly known as GHZ states in quantum information theory literature \cite{pezze2018quantum,hilaire2023graph}). Photonic states generated in superradiant bursts in both waveguide QED and cavity QED systems can also achieve the Heisenberg scaling (Fig. 5i) \cite{paulisch_quantum_2019, abbasgholinejad2025theory}. A key advantage is that, unlike graph-state generation, producing superradiant bursts does not require full coherent control over the quantum emitters \cite{asenjo2017exponential,Sheremet2023,Guerin2016,Das2020,goban2015superradiance,Asaoka2022,McDonnell2022}, and thus may be more straightforward to realize in solid-state QEs. However, the direct usefulness of of the superradiant states to quantum sensing remains an open problem. In particular, the impact of experimental non-idealities as discussed in this review, such as emitter inhomogenieties, photon loss, and dephasing, on the metrological potential of the superradiantly emitted photons still remains to be understood. Moreover, optimal measurement protocols that actually attain the Heisenberg limit with such states need to be identified. The optimal measurement scheme would  require either the generation of optical non-linearities, as theoretically proposed \cite{abbasgholinejad2025theory}, or a case-by-case development of simple observables (such as an optical analogue of spin squeezing parameter) that can be easily measured. 

For optical quantum computing (Fig. 5j), the state-of-the-art strategy involves generating specific cluster states which are computationally universal, i.e., given this resource state, a quantum circuit can be effectively implemented with a specific sequence of measurements and feedback \cite{raussendorf2001one, raussendorf2003measurement}. Furthermore, recent proposals have developed fusion-based quantum computation \cite{bartolucci2023fusion}, where constant-size, independent of the target algorithm graph states (e.g., 4-ring or 6-ring graphs) are generated and consumed as an elementary resource for computation using fusion measurements. However, building an end-to-end quantum computation scheme with QEs, especially while also employing fusion measurements, requires high photon indistinguishability for high fidelities.  Correlated errors are an unavoidable part of photonic graph state generation, as a single QE error can propagate to some of the emitted photons following the error. The correlation time is given by the duration between two measurements of the QE state during the graph state generation protocol. Although the impact of various sources of noise has been considered for fault-tolerance quantum computation with cluster states \cite{bolt2016foliated, paesani2023high, song2024encoded}, it is still a topic of ongoing research how this is maintained under temporally correlated noise.

Graph states have also found an important utility in all-photonic quantum networks \cite{azuma2015all}. This application relies on generating repeater graph states at source nodes before distributing them to intermediate nodes and performing entanglement swapping. Therefore, similarly to fusion-based quantum computation, this application also requires high photon indistinguishability. The generation of repeater graph states with a minimal number of QEs has been discussed in \cite{li2022photonic}. When working with quantum defects in solids, an alternative is to use a single QE and multiple nuclear spins that can be addressed via the defect spin \cite{buterakos2017deterministic}. Improvements in the Debye-Waller factor and advancements in the nuclear spin control pulses for these defects will be important milestones in the generation of repeater-type graph states. Another important proposal that is well-suited for solid-state emitters is given by \cite{borregaard2020one}.

Finally, in addition to being employed as light sources, the system of QEs coupling to a nanophotonic structure can also be a test-bed for simulating many-body physics (Fig. 5k) \cite{noh_quantum_2016, Kennes2021}. The ability to tune the electromagnetic green's function with nanophotonic structures opens up the possibility of engineering long-range coherent and dissipative interaction between quantum emitters (outlined in Box 2). These systems can thus potentially be used as quantum simulators for studying the physics of all-to-all models that arise routinely in quantum optics, condensed matter physics as well as quantum chemistry.  

To realize these potential applications in quantum sensing, computation, communication, and simulation, further advances in solid-state QEs are essential. For more established platforms such as self-assembled III–V quantum dots, color centers in diamond, and DBT molecules in organic crystals, the next key step is to scale coherent control beyond the few-emitter limit. Nanophotonic integration, emitter tuning, active stabilization against spectral diffusion, and high-efficiency collection offer routes to achieving this control. For emitters hosted in 2D-materials, parallel progress must come from improved material growth and a deeper microscopic understanding of coherence. In TMDs and and their heterostructures, strong exciton–exciton interactions and moiré engineering make this platform especially promising for optical nonlinearities, but pushing toward the single-photon nonlinearity regime will require higher cooperativity and longer exciton coherence. In hBN, the open challenge is to achieve simultaneous spin and optical coherence so that a single defect can serve as both a high-quality spin qubit and a source of indistinguishable photons for entanglement generation. For colloidal QDs, encouraging demonstrations of many-body coherence in large ensembles have emerged, but deterministic integration with nanophotonics and individual emitter control are still at an earlier stage of development. Meeting these targets would unlock many-body entanglement from large QE arrays in the solid-state, with direct impact on quantum technologies.

\section*{Acknowledgment}
We acknowledge support from the US Department of Energy, Office of Basic Energy Sciences, through the Quantum Photonic Integrated Design Center (QuPIDC) EFRC award DE-SC0025620.

%%%%%%%%%%%%%%%%%%%%%%%%%%%%%%%%%%%%%%%%%%%%%%%
% =============== .BIB FILE CODE ==============
%%%%%%%%%%%%%%%%%%%%%%%%%%%%%%%%%%%%%%%%%%%%%%%

% ===== UNCOMMENT THESE COMMANDS AND DELETE THE 1000 LINES OF .BBL CODE TO USE THE .BIB FILE =====
% \bibliographystyle{naturemag-doi}
% \bibliography{references_master} 

%%%%%%%%%%%%%%%%%%%%%%%%%%%%%%%%%%%%%%%%%%%%%%%
% ====== .BBL FILE CODE FOR SUBMISSION ========
%%%%%%%%%%%%%%%%%%%%%%%%%%%%%%%%%%%%%%%%%%%%%%%

\end{document}